\documentclass[aps,showpacs,preprintnumbers,amsmath,amssymb,twocolumn,superscriptaddress,prl,floatfix,nofootinbib]{revtex4-1}

\usepackage{amsfonts}
\usepackage{amsmath}
\usepackage{amssymb}
\usepackage{amstext}
\usepackage{amsthm}
\usepackage{graphicx}
\usepackage{dcolumn}
\usepackage{bm}
\usepackage{epsf}
\usepackage{framed}
\usepackage{hyperref}
\usepackage{xcolor}
\usepackage{dsfont}
\usepackage{algpseudocode,algorithm}
\usepackage{mathrsfs}
\usepackage{soul}

\DeclareMathOperator*{\argmax}{arg\,max}
\DeclareMathOperator*{\argmin}{arg\,min}

\begin{document}

\title{Fast and scalable likelihood maximization\\for Exponential Random Graph Models with local constraints}

\author{Nicol\`o Vallarano}
\affiliation{IMT School for Advanced Studies Lucca, P.zza San Francesco 19, 55100 Lucca (Italy)}
\author{Matteo Bruno}
\affiliation{IMT School for Advanced Studies Lucca, P.zza San Francesco 19, 55100 Lucca (Italy)}
\author{Emiliano Marchese}
\affiliation{IMT School for Advanced Studies Lucca, P.zza San Francesco 19, 55100 Lucca (Italy)}
\author{Giuseppe Trapani}
\affiliation{IMT School for Advanced Studies Lucca, P.zza San Francesco 19, 55100 Lucca (Italy)}
\author{Fabio Saracco}
\affiliation{IMT School for Advanced Studies Lucca, P.zza San Francesco 19, 55100 Lucca (Italy)}
\author{Giulio Cimini}
\affiliation{Physics Department and INFN, `Tor Vergata' University of Rome, 00133 Rome (Italy)}
\author{Mario Zanon}
\affiliation{IMT School for Advanced Studies Lucca, P.zza San Francesco 19, 55100 Lucca (Italy)}
\author{Tiziano Squartini}
\email{tiziano.squartini@imtlucca.it}
\affiliation{IMT School for Advanced Studies Lucca, P.zza San Francesco 19, 55100 Lucca (Italy)}
\affiliation{Institute for Advanced Study (IAS), University of Amsterdam, Oude Turfmarkt 145, 1012 GC Amsterdam (The Netherlands)}

\date{\today}

\begin{abstract}
\noindent\textbf{Abstract.} Exponential Random Graph Models (ERGMs) have gained increasing popularity over the years. Rooted into statistical physics, the ERGMs framework has been successfully employed for reconstructing networks, detecting statistically significant patterns in graphs, counting networked configurations with given properties. From a technical point of view, the ERGMs workflow is defined by two subsequent optimization steps: the first one concerns the maximization of Shannon entropy and leads to identify the functional form of the ensemble probability distribution that is maximally non-committal with respect to the missing information; the second one concerns the maximization of the likelihood function induced by this probability distribution and leads to its numerical determination. This second step translates into the resolution of a system of $O(N)$ non-linear, coupled equations (with $N$ being the total number of nodes of the network under analysis), a problem that is affected by three main issues, i.e. \emph{accuracy}, \emph{speed} and \emph{scalability}. The present paper aims at addressing these problems by comparing the performance of three algorithms (i.e. Newton's method, a quasi-Newton method and a recently-proposed fixed-point recipe) in solving several ERGMs, defined by binary and weighted constraints in both a directed and an undirected fashion. While Newton's method performs best for relatively little networks, the fixed-point recipe is to be preferred when large configurations are considered, as it ensures convergence to the solution within seconds for networks with hundreds of thousands of nodes (e.g. the Internet, Bitcoin). We attach to the paper a Python code implementing the three aforementioned algorithms on all the ERGMs considered in the present work.
\end{abstract}


\maketitle

\section{Introduction}\label{introduction}

Over the last 20 years, network theory has emerged as a successful framework to address problems of scientific and societal relevance \cite{Newman2010}: examples of processes that are affected by the structural details of the underlying network are provided by the spreading of infectious diseases \cite{Colizza2006,Barrat2008,Pastor2015}, opinion dynamics \cite{Castellano2009}, the propagation of losses during financial crises \cite{Squartini2013}, etc.

Within such a context, two needs have emerged quite naturally \cite{Cimini2019}: 1) detecting the topological properties of a real network that can be deemed as statistically significant - typically those higher-order properties (e.g. the assortativity, the clustering coefficient, etc.) that may be explained by local features of the nodes such as the degrees, 2) inferring the relevant details of a given network structure in case only partial information is available. Both goals can be achieved by constructing a framework for defining \emph{benchmarks}, i.e. synthetic configurations retaining only some of the properties of the original system - the so-called \emph{constraints} - and, otherwise, being maximally random.

Two different kinds of approaches have been proposed so far, i.e. the \emph{microcanonical} and the \emph{canonical} ones. Microcanonical approaches \cite{MS2002,Coolen2009,Coolen2005,Artzy2005,DelGenio2010,Kim2012,Blitzstein2011} artificially generate many randomized variants of the observed network by enforcing the constraints in a `hard' fashion, i.e. by creating configurations on each of which the constrained properties are identical to the empirical ones. On the other hand, \emph{canonical} approaches \cite{Squartini2011,Newman2004,Bianconi2007,Fronczak2006,Gabrielli2019} enforce constraints in a `soft' fashion, by creating a set of configurations over which the constrained properties are \emph{on average} identical to the empirical ones. Softening the requirement of matching the constraints has a clear advantage: allowing the mathematical expression for the probability of a generic configuration, $\mathbf{G}$, to be obtained analytically, as a function of the enforced constraints.

In this second case, a pivotal role is played by the formalism of the Exponential Random Graph Models (ERGMs) \cite{Fronczak2012} whose popularity has steadily increased over the years. The ERGMs mathematical framework dates back to Gibbs' (re)formulation of statistical mechanics and is based upon the variational principle known as \emph{maximum entropy}, stating that the probability distribution that is maximally non-committal with respect to the missing information is the one maximizing the \emph{Shannon entropy} \cite{Jaynes1957}. This allows self-consistent inference to be made, by assuming maximal ignorance about the unknown degrees of freedom of the system.

In the context of network theory, the ERGMs framework has found a natural application to the resolution of the two aforementioned problems, i.e. 1) that of defining \emph{null models} of the original network, in order to assess the significance of empirical patterns - against the hypothesis that the network structure is solely determined by the constraints themselves and 2) that of deriving the most probable configurations that are compatible with the available details about a specific network.

In both cases, after the functional form of the probability distribution has been identified, via the maximum entropy principle, one also needs to numerically determine it: to this aim, the \emph{likelihood maximization} principle can be invoked, stating to maximize the probability of observing the actual configuration. This prescription typically translates into the resolution of a system of $O(N)$ non-linear, coupled equations - with $N$ representing the number of nodes of the network under analysis.

Problems like these are usually affected by the issues of \emph{accuracy}, \emph{speed} and \emph{scalability}: the present paper aims at addressing them at once, by comparing the performance of three algorithms, i.e. Newton's method, a quasi-Newton method and a recently-proposed fixed-point recipe \cite{Dianati2016,Vallarano2020}, to solve a variety of ERGMs, defined by binary and weighted constraints in both a directed and an undirected fashion.

\textcolor{black}{We would like to stress that, while the theoretical architecture of ERGMs is well established, an exhaustive study of the computational cost required for their numerical optimization, alongside with a comparison among the most popular algorithms designed for the task, is still missing. Our work aims at filling precisely this gap.} Additionally, we provide a Python code implementing the three aforementioned recipes on all the ERGMs considered in the present work.

\section{General theory}\label{methodsI}

Canonical approaches aim at obtaining the mathematical expression for the probability of a generic configuration, $\mathbf{G}$, as a function of the observed constraints: ERGMs realize this by maximizing the Shannon entropy \cite{Squartini2011,Newman2004}.

\subsection{The Maximum Entropy Principle}

Generally speaking, the problem to be solved in order to find the functional form of the probability distribution to be employed as a benchmark reads

\begin{subequations}
\label{eq:primal}
\begin{align}
\argmax_{P} \ \ & S[P] \\
\mathrm{s.t.} \ \ & \sum_{\mathbf{G}}P(\mathbf{G})C_i(\mathbf{G})=\langle C_i\rangle,\quad i=0\dots M
\end{align}
\end{subequations}
where Shannon entropy reads

\begin{equation}
S[P]=-\sum_\mathbf{G}P(\mathbf{G})\ln P(\mathbf{G})
\end{equation}
and $\vec{C}(\mathbf{G})$ is the vector of constraints representing the information defining the benchmark itself (notice that $C_0=\langle C_0\rangle=1$ sums up the normalization condition). The solution to the problem above can be found by maximizing the \emph{Lagrangian function}

\begin{align}
\mathcal{L}(P,\vec{\theta})\equiv S[P]+\sum_{i=0}^M\theta_i\left[-\sum_{\mathbf{G}}P(\mathbf{G})C_i(\mathbf{G})+\langle C_i\rangle\right]
\end{align}
with respect to $P(\mathbf{G})$. As a result one obtains

\begin{equation}\label{eq:p}
P(\mathbf{G}|\vec{\theta})=\frac{e^{-\mathcal{H}(\mathbf{G},\vec{\theta})}}{Z(\vec{\theta})}
\end{equation}
with $\mathcal{H}(\mathbf{G},\vec{\theta})=\vec{\theta}\cdot\vec{C}(\mathbf{G})=\sum_{i=1}^M\theta_iC_i(\mathbf{G})$ representing the \emph{Hamiltonian}, i.e. the function summing up the proper, imposed constraints and $Z(\vec{\theta})=\sum_\mathbf{G}P(\mathbf{G})=\sum_\mathbf{G}e^{-\mathcal{H}(\mathbf{G},\vec{\theta})}$ representing the \emph{partition function}, ensuring that $P(\mathbf{G})$ is properly normalized. Constraints play a pivotal role, either representing the information to filter, in order to assess the significance of certain quantities, or the only available one, in order to reconstruct the inaccessible details of a given configuration.

\begin{figure*}[t!]
\includegraphics[width=\textwidth]{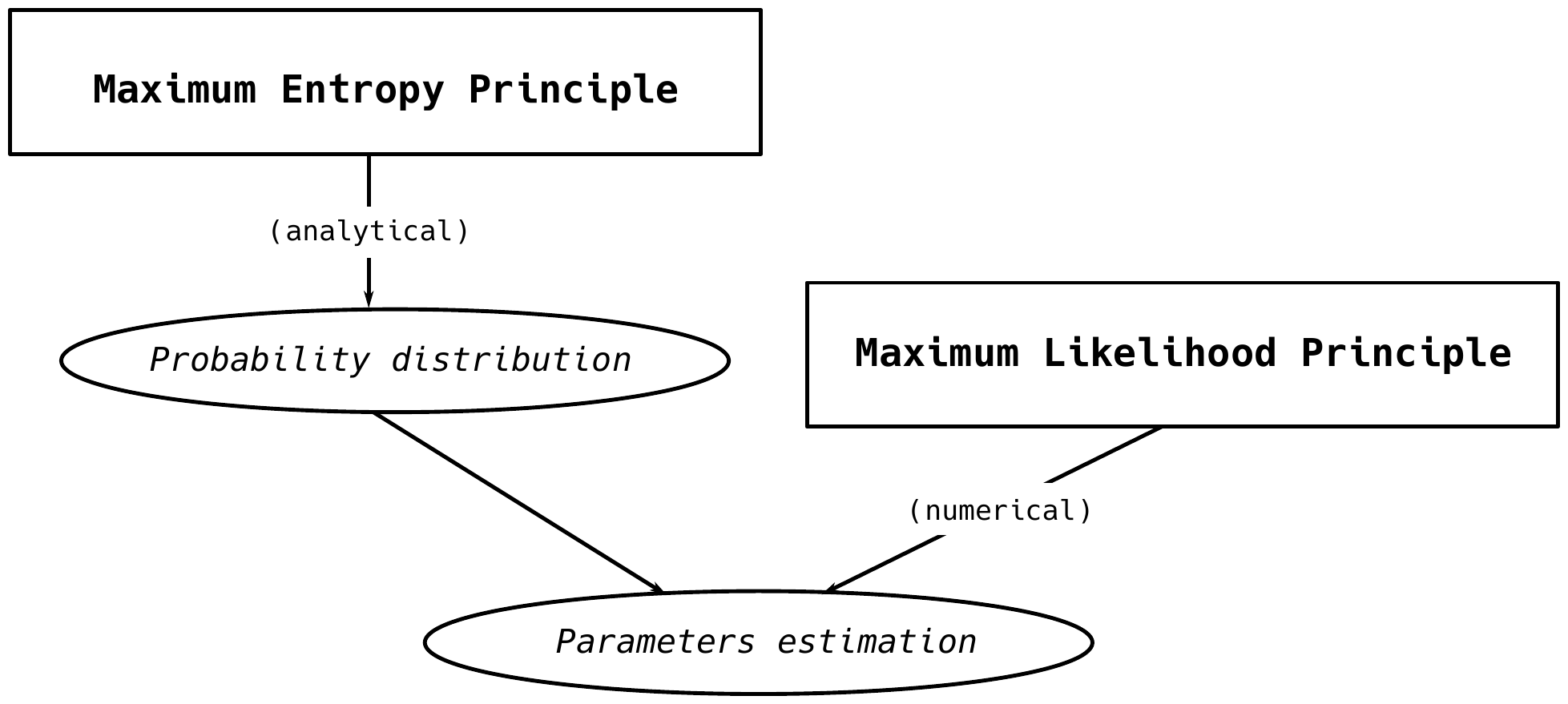}
\caption{\textcolor{black}{Graphical visualization of how the MEP and the MLP work: while the MEP allows the functional form of a probability distribution to be determined analytically, the MLP provides the recipe to numerically determine the parameters defining it.}}
\label{fig1}
\end{figure*}

\subsection{The Maximum Likelihood Principle}

The formalism above is perfectly general; however, it can be instantiated to study an empirical network configuration, say $\mathbf{G}^*$. In this case, the Lagrange multipliers `acting' as unknown parameters in eq. \eqref{eq:p} can be numerically estimated by maximizing the associated likelihood function \cite{Garlaschelli2008,Squartini2011}. The latter is defined as

\begin{equation}\label{eq:l}
\mathscr{L}(\vec{\theta})\equiv\ln P(\mathbf{G}^*|\vec{\theta})=-\mathcal{H}(\mathbf{G}^*,\vec{\theta})-\ln Z(\vec{\theta})
\end{equation}
and must be maximized with respect to the vector $\vec{\theta}$. Remarkably, whenever the probability distribution is exponential (as the one deriving from the Shannon entropy maximization), the likelihood maximization problem

\begin{align}
\argmax_{\vec{\theta}} \ & \mathscr{L}(\vec{\theta})
\end{align}
is characterized by first-order necessary conditions for optimality reading

\begin{eqnarray}
\nabla_{\theta_i}\mathscr{L}(\vec{\theta})=\frac{\partial\mathscr{L}(\vec{\theta})}{\partial\theta_i}&=&-C_i(\mathbf{G}^*)-\frac{\partial\ln Z(\vec{\theta})}{\partial\theta_i}\nonumber\\
&=&-C_i(\mathbf{G}^*)+\sum_\mathbf{G}C_i(\mathbf{G})P(\mathbf{G})\nonumber\\
&=&-C_i(\mathbf{G}^*)+\langle C_i\rangle=0,\quad i=1\dots M\nonumber\\
\end{eqnarray}
and leading to the system of equations

\begin{equation}
\nabla\mathscr{L}(\vec{\theta})=\vec{0}\Longrightarrow\vec{C}(\mathbf{G}^*)=\langle\vec{C}\rangle
\end{equation}
to be solved. These conditions, however, are sufficient to characterize a maximum only if $\mathscr{L}(\vec{\theta})$ is concave. This is indeed the case, as we prove by noticing that

\begin{eqnarray}
H_{ij}&=&\nabla^2_{\theta_i\theta_j}\mathscr{L}(\vec{\theta})=\frac{\partial^2\mathscr{L}(\vec{\theta})}{\partial\theta_i\partial\theta_j}=-\frac{\partial^2\ln Z(\vec{\theta})}{\partial\theta_i\partial\theta_j}\nonumber\\
&=&\frac{\partial\langle C_j\rangle}{\partial\theta_i}=-\text{Cov}[C_i,C_j],\quad i,j=1\dots M
\end{eqnarray}
i.e. that the Hessian matrix, $\mathbf{H}$, of our likelihood function is `minus' the covariance matrix of the constraints, hence negative semidefinite by definition. The fourth passage is an example of the well-known \emph{fluctuation-response relation} \cite{Fronczak2012}.

\textcolor{black}{A graphical representation of how the two principles work is shown in fig. \ref{fig1}.}

\begin{figure*}[t!]
\includegraphics[width=\textwidth]{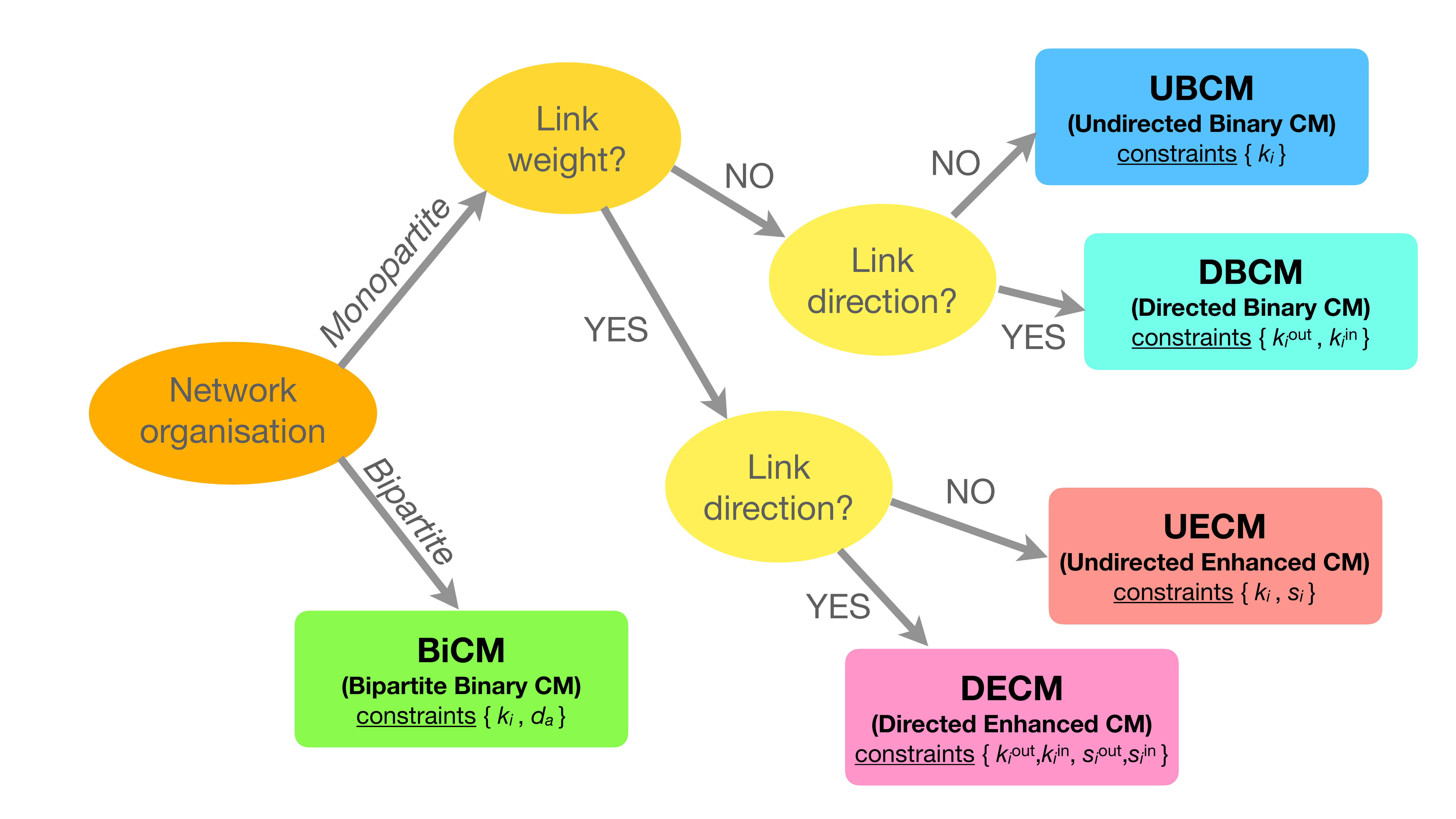}
\caption{\textcolor{black}{System diagram illustrating the models discussed in the present paper and implemented in the `NEMTROPY' package, attached to it: it represents a sort of guide to individuate the best model for analysing the system at hand. Our package handles both monopartite and bipartite networks; while the latter ones have been considered only in their binary, undirected fashion, the former ones can be modeled either in a binary or a weighted fashion, allowing for both undirected and directed links.}}
\label{fig2}
\end{figure*}

\subsection{Combining the MEP and the MLP principles}

\textcolor{black}{The Maximum Entropy Principle (MEP) and the Maximum Likelihood Principle (MLP)} encode two different prescriptions aiming, respectively, at determining the functional form of a probability distribution and its numerical value. In optimization theory, the problem \eqref{eq:primal} is known as \emph{primal problem}: upon noticing that the Shannon entropy is concave, while the imposed constraints are linear in $P(\mathbf{G})$, one concludes that the primal problem is convex (it is easy to see this, by rewriting it as a minimization problem for $-S[P]$).

As convexity implies \emph{strong duality}, we can, equivalently, consider an alternative version of the problem to optimize, know as \emph{dual problem}. In order to define it, let us consider the Lagrangian function

\begin{align}\label{eq:lag}
\mathcal{L}(P,\vec{\theta})\equiv S[P]+\sum_{i=1}^M\theta_i\left[-\sum_{\mathbf{G}}P(\mathbf{G})C_i(\mathbf{G})+C_i(\mathbf{G}^*)\right]
\end{align}
where, now, the generic expectation of the $i$-th constraint, $\langle C_i\rangle$, has been replaced by the corresponding empirical value, $C_i(\mathbf{G}^*)$. As the dual function is given by

\begin{align}
P(\mathbf{G}^*|\vec{\theta})\equiv \argmax_P \ & \mathcal{L}(P,\vec{\theta}),
\end{align}
the dual problem reads

\begin{align}\label{eq:tot}
\argmax_{\vec{\theta}} \ \argmin_P \ -\mathcal{L}(P(\vec{\theta}),\vec{\theta})
\end{align}
which is a convex problem by construction; this is readily seen by substituting eq. \eqref{eq:p} into eq. \eqref{eq:lag}, operation that leads to the expression

\begin{align}
-\mathcal{L}(P(\vec{\theta}),\vec{\theta})=-\vec{\theta}\cdot\vec{C}(\mathbf{G}^*)-\ln Z(\vec{\theta})=\mathscr{L}(\vec{\theta}),
\end{align}
i.e. the likelihood function introduced in eq. \eqref{eq:l}. In other words, eq. \eqref{eq:tot} combines the MEP and the MLP into a unique optimization step whose score function becomes the Lagrangian function defined in eq. \eqref{eq:lag}.

\begin{figure*}[t!]
\includegraphics[width=\textwidth]{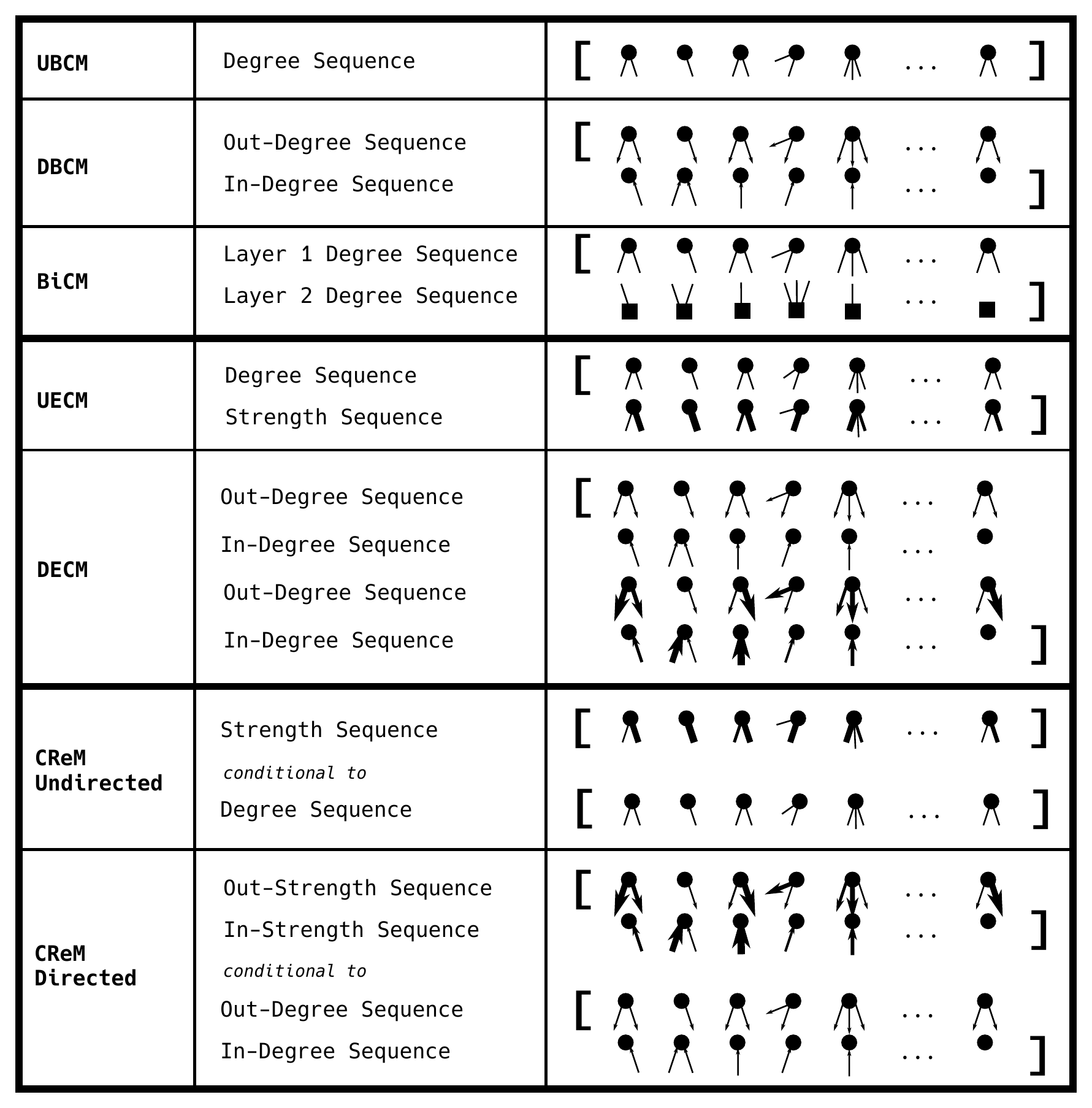}
\caption{\textcolor{black}{Graphical visualization of the constraints defining the ERGMs considered in this work. Notice that while the enhanced models (i.e. UECM and DECM) constrain binary and weighted quantities in a joint fashion, the conditional models (i.e. the CReM ones) allow for a `separate' specification of them.}}
\label{fig3}
\end{figure*}

\subsection{Optimization algorithms for non-linear problems}

In general, the optimization problem defined in eq. \eqref{eq:tot} cannot be solved analytically, whence the need to resort to numerical methods. For an exhaustive review on numerical methods for optimization we refer the interested reader to \cite{Nocedal2006,Boyd2004}: in the following, we present only the concepts that are of some relevance for us. The problem 

\begin{align}\label{eq:nlp}
\argmax_{\vec{\theta}} \ & \mathscr{L}(\vec{\theta})
\end{align}
is a Nonlinear Programming Problem (NLP). In order to solve it numerically, we will adopt a Sequential Quadratic Programming (SQP) approach. Starting from an initial guess $\vec{\theta}^{(0)}$, SQP solves eq. \eqref{eq:nlp} by iteratively updating the vector of Lagrange multipliers

\begin{align}
\vec{\theta}^{(n+1)}_i=\vec{\theta}^{(n)}_i+\alpha\Delta\vec{\theta}^{(n)}_i,\quad i=1\dots M
\end{align}
according to the rule

\begin{equation}
\Delta\vec{\theta}^{(n)}_i=\argmax_{\Delta\vec{\theta}_i} \ \left[\nabla_{\vec{\theta}_i}\mathscr{L}(\vec{\theta})\Delta\vec{\theta}_i+\sum_{j,k}\frac{1}{2}\Delta\vec{\theta}_jH_{jk}^{(n)} \Delta\vec{\theta}_k\right]
\end{equation}
$\forall\:i$, leading to the set of equations

\begin{equation}
\nabla_i\mathscr{L}(\vec{\theta})+\sum_jH_{ij}^{(n)}\Delta\vec{\theta}=0,\quad i=1\dots M
\end{equation}
which can be compactly rewritten as

\begin{equation}\label{eq:gensol}
\Delta\vec{\theta}^{(n)}=-{\mathbf{H}^{(n)}}^{-1}\nabla\mathscr{L}(\vec{\theta}).
\end{equation}

The stepsize $\alpha\in(0,1]$ is selected to ensure that $\mathscr{L}(\vec{\theta}^{(n+1)})>\mathscr{L}(\vec{\theta}^{(n)})$ via a backtracking, line search procedure: starting from $\alpha=1$, if the Armijo condition

\begin{align}
\mathscr{L}(\vec{\theta}^{(n)}+\alpha\Delta \vec{\theta}^{(n)})<\mathscr{L}(\vec{\theta}^{(n)})+\gamma \alpha \nabla \mathscr{L}(\vec{\theta})^\top \Delta \vec{\theta},
\end{align}
is violated, we set $\alpha \leftarrow \beta \alpha$ ($\gamma\in(0,0.5]$ and $\beta\in(0,1)$ are the parameters of the algorithm). On the other hand, the term $\mathbf{H}^{(n)}$ can be selected according to a variety of methods. In the present contribution we focus on the following three ones.\\

\paragraph{Newton's method.} One speaks of Newton's method in case $\mathbf{H}^{(n)}$ is chosen to be

\begin{align}
\label{eq:exact_hessian}
\mathbf{H}^{(n)}=\nabla^2\mathscr{L}(\vec{\theta}^{(n)})+\Delta\mathbf{H}^{(n)}
\end{align}
where $\nabla^2\mathscr{L}(\vec{\theta})$ is the Hessian matrix of the likelihood function and the term $\Delta\mathbf{H}^{(n)}$ is typically selected as small as possible in order to avoid slowing convergence and to ensure that $\mathbf{H}^{(n)}$ is negative definite (i.e. $\nabla^2\mathscr{L}(\vec{\theta}^{(n)})+\Delta\mathbf{H}^{(n)}\prec 0$). This choice of $\mathbf{H}^{(n)}$ is also referred to as `exact Hessian'.\\

\paragraph{Quasi-Newton methods.} Any Hessian approximation which is negative definite (i.e. satisfying $\mathbf{H}^{(n)}\prec 0$) yields an ascent direction and guarantees convergence. Although one may choose to consider the simplest prescription $\mathbf{H}^{(n)}=-\mathbf{I}$, which yields the `steepest ascent' algorithm, here we have opted for the following recipe, i.e. the purely diagonal version of Newton's method: $H_{ii}^{(n)}=\nabla^2_{ii}\mathscr{L}(\vec{\theta}^{(n)})+\Delta H_{ii}^{(n)}<0$, $\forall\:i$ and $H_{ij}^{(n)}=0$, $\forall\:i\neq j$.\\

\paragraph{Fixed-point iteration on modified KKT conditions.} In addition to the (classes of) algorithms above, we will also consider an iterative recipe which is constructed as a fixed-point iteration on a modified version of the Karush–Kuhn–Tucker (KKT) conditions, i.e. $\mathbf{F}(\vec{\theta})=\vec{0}$ or, analogously, $\vec{\theta}=\mathbf{G}(\vec{\theta})$; the iterate can, then, be made explicit by rewriting the latter as

\begin{align}\label{eq:fp}
\vec{\theta}^{(n)}=\mathbf{G}(\vec{\theta}^{(n-1)}).
\end{align}

The condition above will be made explicit, for each network model, in the corresponding subsection. We also observe that this choice yields a non-standard SQP method as $\mathbf{H}^{(n)}$ is typically not symmetric, for our models.\\

\textcolor{black}{\paragraph{A note on convergence.} Provided that the Hessian approximation $\mathbf{H}^{(n)}$ is negative definite, the direction $\Delta\vec{\theta} $ is an ascent one; as such, it is guaranteed to yield an improvement of the objective function, for a step size $\alpha$ that is sufficiently small. The role of the backtracking line search is that of finding a step size $\alpha$ that yields such an improvement, while making sufficient progress towards the solution. As discussed in \cite{Nocedal2006}, Newton's method has local quadratic convergence, while the quasi-Newton method and the fixed-point iteration algorithm have local linear convergence.}

\section{Applications}\label{methodsII}

Let us now apply the algorithms described in the previous section to a number of specific cases of interest. \textcolor{black}{The taxonomy of the models considered in the present paper is shown in fig. \ref{fig2} while the constraints defining each model are illustrated in fig. \ref{fig3}.}

\subsection{UBCM: binary undirected graphs\\with given degree sequence\label{sec:UBCM}}

Let us start by considering binary, undirected networks (BUNs). The simplest, non-trivial set of constraints is represented by the degrees of nodes: the degree of node $i$, i.e. $k_i(\mathbf{A})=\sum_{j(\neq i)=1}^N a_{ij}$, counts the number of its neighbours and coincides with the total number of 1s along the $i$-th row (or, equivalently, along the $i$-th column) of the adjacency matrix $\mathbf{A}\equiv\{a_{ij}\}_{i,j=1}^N$. The benchmark defined by this set of constraints is known as \emph{Undirected Binary Configuration Model} (UBCM) and its Hamiltonian reads

\begin{equation}
\mathcal{H}_\text{UBCM}(\mathbf{A},\vec{\theta})=\sum_{i=1}^N\theta_i k_i(\mathbf{A});
\end{equation}
entropy maximization \cite{Newman2004,Squartini2011} leads to the factorized graph probability

\begin{equation}
P_\text{UBCM}(\mathbf{A}|\vec{\theta})=\prod_{i=1}^N\prod_{\substack{j=1\\(j<i)}}^Np_{ij}^{a_{ij}}(1-p_{ij})^{1-a_{ij}}\label{cm}
\end{equation}
where $p_{ij}=p_{ij}^\text{UBCM}\equiv\frac{e^{-\theta_i-\theta_j}}{1+e^{-\theta_i-\theta_j}}$. In this case, the canonical ensemble of BUNs is the set of networks with the same number of nodes, $N$, of the observed graph and a number of (undirected) links varying from zero to the maximum value $\binom{N}{2}$. The argument of the problem \eqref{eq:nlp} for the specific network $\mathbf{A}^*$ becomes

\begin{equation}
\mathscr{L}_\mathrm{UBCM}(\vec{\theta})=-\sum_{i=1}^N\theta_ik_i(\mathbf{A}^*)-\sum_{i=1}^N\sum_{\substack{j=1\\(j<i)}}^N\ln\left[1+e^{-\theta_i-\theta_j}\right]
\end{equation}
whose first-order optimality conditions read

\begin{eqnarray}\label{eq:sysubcm}
\nabla_{\theta_i}\mathscr{L}_\mathrm{UBCM}&=&-k_i(\mathbf{A}^*)+\sum_{\substack{j=1\\(j\neq i)}}^N\frac{e^{-\theta_i-\theta_j}}{1+e^{-\theta_i-\theta_j}}\nonumber\\
&=&-k_i(\mathbf{A}^*)+\sum_{\substack{j=1\\(j\neq i)}}^Np_{ij}^\text{UBCM}\nonumber\\
&=&-k_i(\mathbf{A}^*)+\langle k_i\rangle=0,\quad i=1\dots N.\nonumber\\
\end{eqnarray}

\paragraph{Resolution of the UBCM.} Newton's and the quasi-Newton method can be easily implemented via the recipe defined in eq. \eqref{eq:gensol} (see Appendix A for the definition of the UBCM Hessian).

The explicit definition of the fixed-point recipe, instead, requires a preliminary observation, i.e. that the system of equations embodying the UBCM first-order optimality conditions can be re-written as follows

\begin{equation}\label{eq:sys0}
e^{-\theta_i}=\frac{k_i(\mathbf{A}^*)}{\sum_{\substack{j=1\\(j\neq i)}}^N\left(\frac{e^{-\theta_j}}{1+e^{-\theta_i-\theta_j}}\right)},\quad i=1\dots N
\end{equation}
i.e. as a set of consistency equations. The observation that the term $e^{-\theta_i}$ appears on both sides of the equation corresponding to the $i$-th constraint suggests an iterative recipe to solve such a system, i.e.

\begin{equation}\label{eq:sys}
\theta_i^{(n)}=-\ln\left[\frac{k_i(\mathbf{A}^*)}{\sum_{\substack{j=1\\(j\neq i)}}^N\left(\frac{e^{-\theta_j^{(n-1)}}}{1+e^{-\theta_i^{(n-1)}-\theta_j^{(n-1)}}}\right)}\right],\quad i=1\dots N
\end{equation}
originally proposed in \citep{Dianati2016} and further refined in \citep{Vallarano2020}. The identification $p_{ij}^\text{UBCM}\equiv\frac{e^{-\theta_i^{(\infty)}-\theta_j^{(\infty)}}}{1+e^{-\theta_i^{(\infty)}-\theta_j^{(\infty)}}}$, $\forall\:i<j$ allows the probability coefficients defining the UBCM to be numerically determined.

As any other iterative recipe, the one proposed above needs to be initialized as well. To this aim, we have tested three different sets of initial values: the first one is defined by the position $\theta_i^{(0)}=-\ln\left[\frac{k_i(\mathbf{A}^*)}{\sqrt{2L}}\right]$, $\forall\:i$ - usually, a good approximation of the solution of the system of equations in \eqref{eq:sysubcm}, in the `sparse case' (i.e. whenever $p_{ij}^\text{UBCM}\simeq e^{-\theta_i-\theta_j}$ \cite{Chung2002}; the second one is a variant of the position above, reading $\theta_i^{(0)}=-\ln\left[\frac{k_i(\mathbf{A}^*)}{\sqrt{N}}\right]$, $\forall\:i$; the third one, instead, prescribes to randomly draw the value of each parameter from a uniform distribution with support on the unit interval, i.e. $\theta_i^{(0)}\sim U(0,1)$, $\forall\:i$.\\

\begin{table*}[t!]
\centering
\begin{tabular}{|l|c|c|c|c||c|c||c|c||c|c|}
\hline
\multicolumn{1}{c}{} & \multicolumn{1}{c}{} & \multicolumn{1}{c}{} & \multicolumn{1}{c}{} & \multicolumn{1}{c}{} & \multicolumn{2}{c}{Newton} & \multicolumn{2}{c}{Quasi-Newton} & \multicolumn{2}{c}{Fixed-point}\\
\hline
& $N$ & $L$ & $c$ & $c_{r}$ & MADE & Time (s) & MADE & Time (s) & MADE & Time (s)\\
\hline
\textit{C. Elegans} (nn) & 265 & 1879 & $\simeq 5\cdot 10^{-2}$ & $\simeq  1.5 \cdot 10^{-1}$ &
$\simeq 8.1 \cdot 10^{-8}$ & $\simeq 0.005$ & 
$\simeq 1 \cdot 10^{-6}$ & $\simeq 0.03$ & 
$\simeq 5\cdot 10^{-8}$ & $\simeq 0.004$\\
\hline
US airports & 500 & 2980 & $\simeq 2\cdot 10^{-2}$ & $\simeq 1.3\cdot 10^{-1}$ &
$\simeq 8.9 \cdot 10^{-9}$ & $\simeq 0.008$ &
$\simeq 1.2 \cdot 10^{-6}$ & $\simeq 0.04 $ &
$\simeq 2.5 \cdot 10^{-7}$ & $\simeq 0.005$\\
\hline
\textit{H. Pylori} (pp) & 732 & 1465 & $\simeq 5\cdot 10^{-3}$ & $\simeq 4.5 \cdot 10^{-2}$ &
$\simeq 1.3 \cdot 10^{-8}$ & $\simeq 0.004 $ &
$\simeq 6.7 \cdot 10^{-7}$ & $\simeq 0.03$ &
$\simeq 7 \cdot 10^{-8}$ & $\simeq 0.014 $\\
\hline
Internet (AS) & 11174 & 23409 & $\simeq 4\cdot 10^{-4}$ & $\simeq  1 \cdot 10^{-2}$ & $\simeq 4.1 \cdot 10^{-7}$ & $\simeq 0.03 $ &
$\simeq 5.1 \cdot 10^{-6}$ & $\simeq 0.10 $ &
$\simeq 2 \cdot 10^{-6}$ & $\simeq 0.005$\\
\hline
\hline
BLN 24-01-18 & 94 & 152 & $\simeq 3 \cdot 10^{-3}$ & $\simeq 1.5 \cdot 10^{-1}$ & $\simeq 1 \cdot 10^{-8}$ & $\simeq 0.005 $ &
$\simeq 1.7 \cdot 10^{-6}$ & $\simeq 0.014$ &
$\simeq 4 \cdot 10^{-8}$ & $\simeq 0.002 $\\
\hline
BLN 25-02-18 & 499 & 1010 & $\simeq 8 \cdot 10^{-3}$ & $\simeq 6.4\cdot 10^{-2}$ & $\simeq 1.6 \cdot 10^{-8}$ & $\simeq 0.005 $ &
$\simeq 9.1 \cdot 10^{-7}$ & $\simeq 0.02 $ &
$\simeq 7.3 \cdot 10^{-8}$ & $\simeq 0.004 $\\
\hline
BLN 30-03-18 & 1012 & 2952 & $\simeq 5 \cdot 10^{-3}$ & $\simeq 5.2\cdot 10^{-2}$ & $\simeq 3.4 \cdot 10^{-10}$ & $\simeq 0.005 $ &
$\simeq 2 \cdot 10^{-5}$ & $\simeq 0.03$ &
$\simeq 1.4 \cdot 10^{-7}$ & $\simeq 0.005$\\
\hline
BLN 13-07-18 & 1999 & 8999 & $\simeq 4 \cdot 10^{-3}$ & $\simeq 3.8\cdot 10^{-2}$ & $\simeq 6.7 \cdot 10^{-10}$ & $\simeq 0.01 $ &
$\simeq 3 \cdot 10^{-6}$ & $\simeq 0.05$ &
$\simeq 1.7 \cdot 10^{-7}$ & $\simeq 0.008 $\\
\hline
BLN 19-12-18 & 3007 & 17689 & $\simeq 4 \cdot 10^{-3}$ & $\simeq 4.5\cdot 10^{-2}$ & $\simeq 1.7 \cdot 10^{-6}$ & $\simeq  0.03 $ &
$\simeq 1 \cdot 10^{-5}$ & $ \simeq 0.09 $ &
$\simeq 2 \cdot 10^{-7}$ & $\simeq 0.010 $\\
\hline
BLN 30-01-19 & 3996 & 27429 & $\simeq 3 \cdot 10^{-3}$ & $\simeq 4.2\cdot 10^{-2}$ & $\simeq 7.2 \cdot 10^{-7}$ & $\simeq 0.04 $ &
$\simeq 1.3 \cdot 10^{-5}$ & $\simeq 0.12$ &
$\simeq 2.7 \cdot 10^{-7}$ & $\simeq 0.012$\\
\hline
BLN 01-03-19 & 5012 & 41096 & $\simeq 3 \cdot 10^{-3}$ & $\simeq 4.7\cdot 10^{-2}$ & $\simeq 3.9 \cdot 10^{-10}$ & $\simeq 0.03$ &
$\simeq 1.7 \cdot 10^{-4}$ & $\simeq 0.13 $ &
$\simeq 2.6 \cdot 10^{-7}$ & $\simeq 0.013$\\
\hline
BLN 17-07-19 & 6447 & 54476 & $\simeq 3 \cdot 10^{-3}$ & $\simeq 3.3\cdot 10^{-2}$ & $\simeq 4.1 \cdot 10^{-6}$ & $\simeq 0.06 $ &
$\simeq 1.2 \cdot 10^{-5}$ & $\simeq 0.15 $ &
$\simeq 3.1 \cdot 10^{-7}$ & $\simeq 0.015$\\
\hline
\end{tabular}
\caption{Performance of Newton's, quasi-Newton and the fixed-point algorithm to solve the reduced system of equations defining the UBCM, on a set of real-world BUNs (of which basic statistics as the total number of nodes, $N$, the total number of links, $L$, and the connectance, $c=2L/N(N-1)$, are provided). All algorithms stop either because the condition $||\nabla\mathscr{L}(\vec{\theta})||_2\leq 10^{-8}$ is satisfied or because the condition $||\Delta\vec{\theta}||_2\leq 10^{-8}$ is satisfied. For what concerns accuracy, the two most accurate methods are Newton's and the fixed-point ones; for what concerns speed, the fastest method is the fixed-point one (although Newton's one approximately requires the same amount of time on each specific configuration). Only the results corresponding to the best choice of initial conditions are reported.}
\label{tab:undirected}
\end{table*}

\paragraph{Reducing the dimensionality of the problem.} The problem defining the UBCM can be further simplified by noticing that nodes with the same degree, say $k$, can be assigned the same value of the multiplier $\theta$ \citep{Garlaschelli2008} - a result resting upon the observation that any value $k_i(\mathbf{A}^*)$ must match the sum of monotonic, increasing functions. This translates into the possibility of rewriting $\mathscr{L}_\mathrm{UBCM}(\vec{\theta})$ in a `reduced' fashion, as

\begin{eqnarray}
\mathscr{L}_\mathrm{UBCM}^\text{reduced}(\vec{\theta})&=&-\sum_kf(k)\theta_kk(\mathbf{A}^*)\nonumber\\
&&-\sum_k\sum_{\substack{k'\\(k'\leq k)}}f(k)[f(k')-\delta_{kk'}]\cdot\nonumber\\
&&\cdot\ln\left[1+e^{-\theta_k-\theta_{k'}}\right]
\end{eqnarray}
where the sums run over the \emph{distinct} values of the degrees and $f(k)$ counts the number of nodes whose degree is $k$. Rewriting the problem with respect to the set $\{\theta_k\}_k$ leads one to recover simplified versions of the three algorithms considered here: Newton's and the quasi-Newton methods can, now, be solved via a `reduced' version of eq. \eqref{eq:gensol} (since both the dimension of the gradient and the order of the Hessian matrix of the likelihood function are, now, less than $N$), while the iterative recipe defined in \eqref{eq:sys0} can be rewritten in terms of the `non-degenerate' degrees, as

\begin{equation}\label{eq:sys2}
\theta_k^{(n)}=-\ln\left[\frac{k(\mathbf{A}^*)}{\sum_{k'}[f(k')-\delta_{kk'}]\left(\frac{e^{-\theta_{k'}^{(n-1)}}}{1+e^{-\theta^{(n-1)}_{k}-\theta^{(n-1)}_{k'}}}\right)
}\right]
\end{equation}
$\forall\:k$, where, at the denominator, the self-contribution (i.e. the probability that a node links to itself) has been explicitly excluded.\\

\paragraph{Performance testing.} The performance of the three algorithms, considered in the present paper, to solve the reduced version of eqs. \eqref{eq:sysubcm}, has been tested on a bunch of real-world networks. The latter ones span a wide variety of systems, including natural, financial and technological ones. In particular, we have considered the synaptic network of the worm \emph{C. Elegans} \citep{Oshio2003}, the network of the largest US airports \citep{Colizza2007}, the protein-protein interaction network of the bacterium \emph{H. Pylori} \citep{HPylori}, Internet at the level of Autonomous Systems \cite{Colizza2006b} and eight daily snapshots of the so-called Bitcoin Lightning Network \citep{Lin2020}, chosen throughout its entire history. Before commenting on the results of our numerical exercises, let us, first, describe how the latter ones have been carried out.\\

The accuracy of each algorithm in reproducing the constraints defining the UBCM has been quantified via the \emph{maximum absolute error} metrics, defined, in a perfectly general fashion, as $\max_i\{\left|C_i^*-\langle C_i \rangle\right|\}_{i=1}^N$ (where $C_i^*$ is the empirical value of the $i$-th constraint, $C_i$). Naturally, in the UBCM case, $C_i=k_i$, $\forall\:i$ and the aforementioned error score becomes

\begin{equation}
\text{MADE}=\max_i\{\left|k_i^*-\langle k_i \rangle\right|\}_{i=1}^N
\end{equation}
(the acronym standing for Maximum Absolute Degree Error). Equivalently, it is the infinite norm of the difference between the vector of the empirical values of the constraints and that of their expected values.

For each algorithm, we have considered three different stopping criteria: the first one puts a condition on the Euclidean norm of the gradient of the likelihood function, i.e.

\begin{equation}\label{eq:stop1}
||\nabla\mathscr{L}(\vec{\theta})||_2=\sqrt{\sum_{i=1}^N\left(\nabla_i\mathscr{L}(\vec{\theta})\right)^2}\leq 10^{-8};
\end{equation}
the second one puts a condition on the Euclidean norm of the vector of differences between the values of the parameters at subsequent iterations, i.e.

\begin{equation}\label{eq:stop2}
||\Delta\vec{\theta}||_2=\sqrt{\sum_{i=1}^N\left(\Delta\theta_i\right)^2}\leq 10^{-8};
\end{equation}
the third one concerns the maximum number of iterations: after 1000 steps, any of the three algorithms stops.\\

The results about the performance of our three algorithms are reported in \textcolor{black}{Table \ref{tab:undirected}}. Overall, all recipes perform very satisfactorily, being accurate, fast and scalable; moreover, all algorithms stop either because the condition on the norm of the likelihood is satisfied or because the condition on the norm of the vector of parameters is satisfied.

For what concerns accuracy, the largest maximum error per method spans an interval (across all configurations) that amounts at $10^{-10}\lesssim\text{MADE}_\text{Newton}^\text{reduced}\lesssim 10^{-6}$, $10^{-6}\leq\text{MADE}_\text{Quasi-Newton}^\text{reduced}\leq 10^{-5}$ and $10^{-8}\lesssim\text{MADE}_\text{fixed-point}^\text{reduced}\lesssim 10^{-6}$. By looking at each specific network, it is evident that the two most accurate methods are systematically Newton's and the fixed-point ones.

For what concerns speed, the amount of time required by each method to achieve convergence spans an interval (across all configurations) that is $0.005\leq T_\text{Newton}^\text{reduced}\leq 0.01$, $0.014\leq T_\text{Quasi-Newton}^\text{reduced}\leq 0.15$ and $0.002\leq T_\text{fixed-point}^\text{reduced}\leq 0.015$ (time is measured in seconds). The fastest method is the fixed-point one, although Newton's method approximately requires the same amount of time, when compared to it on each specific configuration. Differences in the speed of convergence of any method, caused by the choice of a particular set of initial conditions, are indeed observable: the prescription reading $\theta_i^{(0)}=-\ln\left[\frac{k_i(\mathbf{A}^*)}{\sqrt{N}}\right]$, $\forall\:i$ outperforms the other ones.

Let us now comment on the scalability of our algorithms. What we learn from our exercise is that scalability is not related to the network size in a simple way: the factors seemingly playing a major role are the ones affecting the reducibility of the original system of equations, i.e. the ones `deciding' the number of different equations that actually need to be solved.

While reducibility can be easily quantified \textit{a posteriori}, e.g. by calculating the \textit{coefficient of reduction}, $c_r$, defined as the ratio between the number of equations that survive to reduction and the number of equations defining the original problem (hence, the smaller the better), providing an exhaustive list of the aforementioned factors \textit{a priori} is much more difficult.

\textcolor{black}{In the case of the UBCM, $c_r$ is defined as the number of different degrees divided by the total number of nodes;} one may, thus, argue that reducibility is affected by the heterogeneity of the degree distribution; upon considering that the latter can be quantified by computing the \textit{coefficient of variation} (defined as $c_v=s/m$, where $s$ and $m$ are, respectively, the standard deviation and the mean of the degree distribution of the network at hand), one may derive a simple rule of thumb: a larger coefficient of variation (pointing out a larger heterogeneity of the degree distribution) leads to a larger coefficient of reduction and a larger amount of time for convergence will be required. \textcolor{black}{Notice that even if the degree distribution is narrow, outliers (e.g. hubs) may still play a role, forcing the corresponding parameters to assume either very large or very small values - hence, slowing down the entire convergence process.}

In this sense, scalability is the result of a (non-trivial) interplay between size and reducibility. Let us take a look at \textcolor{black}{Table \ref{tab:undirected}}: Internet is the most reducible network of our basket, although being the largest in size, while the neural network of \emph{C. Elegans} is one of the least reducible networks of our basket, although being the second smallest one; as a consequence, the actual number of equations defining the UBCM on \emph{C. Elegans} is $\simeq 30$ while the actual number of equations defining the UBCM on Internet is $\simeq 100$ - whence the larger amount of time to solve the latter. Remarkably, the time required by our recipes to ensure that the largest system of equations converges to the solution ranges from thousandths to tenths of seconds.\\

As a last comment, we would like to stress that, unlike several popular approximations as the Chung-Lu one \cite{Chung2002}, the generic coefficient $p_{ij}^\text{UBCM}$ always represents a proper probability, in turn implying that eq. \eqref{cm} also provides us with a recipe to sample the canonical ensemble of BUNs, under the UBCM. Notice that the factorization of the graph probability $P_\text{UBCM}(\mathbf{A}|\vec{\theta})$ greatly simplifies the entire procedure, allowing a single graph to be sampled by implementing the Bernoulli trial

\begin{equation}
a_{ij}=
\begin{cases}
0 \: & 1-p_{ij}^\text{UBCM} \\
1 \: & p_{ij}^\text{UBCM}
\end{cases}
\end{equation}
for each (undirected) pair of nodes, in either a sequential or a parallel fashion. The sampling process, whose computational complexity amounts at $O(N^2)$, can be repeated to generate as many configurations as desired. The pseudo-code for explicitly sampling the UBCM ensemble is summed up by Algorithm 1.

\textcolor{black}{We explicitly acknowledge the existence of the algorithm proposed in \cite{Miller2011} for sampling binary, undirected networks from the Chung-Lu model (i.e. the `sparse case' approximation of the UBCM), a recipe that is applicable whenever the condition $p_{ij}^\text{CL}=\frac{k_ik_j}{2L}<1$, $\forall\:i<j$ is verified. As explicitly acknowledged by the authors of \cite{Miller2011}, however, it does not hold in several cases of interest: an example of paramount importance is provided by sparse networks whose degree distribution is scale-free. In such cases, $k_{max}\sim N^{\frac{1}{\gamma-1}}$: hence, the hubs establish a connection with probability $p_{ij}^\text{CL}\sim\frac{N^{\frac{2}{\gamma-1}}}{N-N^{\frac{1}{\gamma-1}}}$ that becomes larger than 1 when $2<\gamma\leq 3$ and diverges for $\gamma\rightarrow 2$, thus leading to a strong violation of the requirement above.}

\begin{algorithm}[H]
\caption{Sampling the UBCM ensemble}
\label{code:undirected}
\begin{algorithmic}[1]
\For{$m=1\dots |E|$}
\State $\mathbf{A}=\mathbf{0};$
\For{$i=1\dots N$}
\For{$j=1\dots N$ \textrm{and} $j<i$}
\If{$\text{RandomUniform}[0,1]\leq p_{ij}^\text{UBCM}$}
\State $a_{ij}=a_{ji}=1;$
\Else
\State $a_{ij}=a_{ji}=0;$
\EndIf
\EndFor
\EndFor
\State $\text{Ensemble}[m]=\mathbf{A};$
\EndFor
\end{algorithmic}
\end{algorithm}

\subsection{DBCM: binary directed graphs\\with given in-degree and out-degree sequences\label{sec:DBCM}}

Let us now move to consider binary, directed networks (BDNs). In this case, the simplest, non-trivial set of constraints is represented by the in-degrees and the out-degrees of nodes, where $k_i^{in}(\mathbf{A})=\sum_{j(\neq i)=1}^Na_{ji}$ counts the number of nodes `pointing' to node $i$ and $k_i^{out}(\mathbf{A})=\sum_{j(\neq i)=1}^Na_{ij}$ counts the number of nodes $i$ `points' to. The benchmark defined by this set of constraints is known as \emph{Directed Binary Configuration Model} (DBCM) whose Hamiltonian reads

\begin{equation}
\mathcal{H}_\text{DBCM}(\mathbf{A},\vec{\alpha},\vec{\beta})=\sum_{i=1}^N[\alpha_i k_i^{out}(\mathbf{A})+\beta_i k_i^{in}(\mathbf{A})];
\end{equation}
as in the undirected case, entropy maximization \cite{Newman2004,Squartini2011} leads to a factorized probability distribution, i.e.

\begin{equation}
P_\text{DBCM}(\mathbf{A}|\vec{\alpha},\vec{\beta})=\prod_{i=1}^N\prod_{\substack{j=1\\(j\neq i)}}^Np_{ij}^{a_{ij}}(1-p_{ij})^{1-a_{ij}}\label{dcm}
\end{equation}
where $p_{ij}=p_{ij}^\text{DBCM}\equiv\frac{e^{-\alpha_i-\beta_j}}{1+e^{-\alpha_i-\beta_j}}$. The canonical ensemble of BDNs is, now, the set of networks with the same number of nodes, $N$, of the observed graph and a number of (directed) links varying from zero to the maximum value $N(N-1)$. The argument of the problem \eqref{eq:nlp} for the specific network $\mathbf{A}^*$ becomes

\begin{eqnarray}\label{eq:dbcm}
\mathscr{L}_\text{DBCM}(\vec{\alpha},\vec{\beta})=&-&\sum_{i=1}^N[\alpha_ik_i^{out}(\mathbf{A}^*)+ \beta_ik_i^{in}(\mathbf{A}^*)]\nonumber\\
&-&\sum_{i=1}^N\sum_{\substack{j=1\\(j\neq i)}}^N\ln\left[1+e^{-\alpha_i-\beta_j}\right]
\end{eqnarray}
whose first-order optimality conditions read

\begin{eqnarray}\label{sys:dbcm1}
\nabla_{\alpha_i}\mathscr{L}_\mathrm{DBCM}&=&-k_i^{out}(\mathbf{A}^*)+\sum_{\substack{j=1\\(j\neq i)}}^N\frac{e^{-\alpha_i-\beta_j}}{1+e^{-\alpha_i-\beta_j}}\nonumber\\
&=&-k_i^{out}(\mathbf{A}^*)+\sum_{\substack{j=1\\(j\neq i)}}^Np_{ij}^\text{DBCM}\nonumber\\
&=&-k_i^{out}(\mathbf{A}^*)+\langle k_i^{out}\rangle=0,\quad i=1\dots N\nonumber\\
\end{eqnarray}
and

\begin{eqnarray}\label{sys:dbcm2}
\nabla_{\beta_i}\mathscr{L}_\mathrm{DBCM}&=&-k_i^{in}(\mathbf{A}^*)+\sum_{\substack{j=1\\(j\neq i)}}^N\frac{e^{-\alpha_j-\beta_i}}{1+e^{-\alpha_j-\beta_i}}\nonumber\\
&=&-k_i^{in}(\mathbf{A}^*)+\sum_{\substack{j=1\\(j\neq i)}}^Np_{ji}^\text{DBCM}\nonumber\\
&=&-k_i^{in}(\mathbf{A}^*)+\langle k_i^{in}\rangle=0,\quad i=1\dots N.\nonumber\\
\end{eqnarray}

\paragraph{Resolution of the DBCM.} Newton's and the quasi-Newton method can be easily implemented via the recipe defined in eq. \eqref{eq:gensol} (see Appendix A for the definition of the DBCM Hessian).

The fixed-point recipe for solving the system of equations embodying the DBCM first-order optimality conditions can, instead, be re-written in the usual iterative fashion as follows

\begin{eqnarray}
\alpha_i^{(n)}=-\ln\left[\frac{k_i^{out}(\mathbf{A}^*)}{\sum_{\substack{j=1\\(j\neq i)}}^N\left(\frac{e^{-\beta_j^{(n-1)}}}{1+e^{-\alpha_i^{(n-1)}-\beta_j^{(n-1)}}}\right)}\right],\quad i=1\dots N\nonumber\\
\beta_i^{(n)}=-\ln\left[\frac{k_i^{in}(\mathbf{A}^*)}{\sum_{\substack{j=1\\(j\neq i)}}^N\left(\frac{e^{-\alpha_j^{(n-1)}}}{1+e^{-\alpha_j^{(n-1)}-\beta_i^{(n-1)}}}\right)}\right],\quad i=1\dots N\nonumber\\
\end{eqnarray}

Analogously to the undirected case, the initialization of this recipe has been implemented in three different ways. The first one reads $\alpha_i^{(0)}=-\ln\left[\frac{k_i^{out}(\mathbf{A}^*)}{\sqrt{L}}\right]$, $i=1\dots N$ and $\beta_i^{(0)}=-\ln\left[\frac{k_i^{in}(\mathbf{A}^*)}{\sqrt{L}}\right]$, $i=1\dots N$ and represents a good approximation to the solution of the system of equations defining the DBCM in the `sparse case' (i.e. whenever $p_{ij}^\text{DBCM}\simeq e^{-\alpha_i-\beta_j}$); the second one is a variant of the position above, reading $\alpha_i^{(0)}=-\ln\left[\frac{k_i^{out}(\mathbf{A}^*)}{\sqrt{N}}\right]$, $i=1\dots N$ and $\beta_i^{(0)}=-\ln\left[\frac{k_i^{in}(\mathbf{A}^*)}{\sqrt{N}}\right]$, $i=1\dots N$; the third one, instead, prescribes to randomly draw the value of each parameter from a uniform distribution defined on the unit interval, i.e. $\alpha_i^{(0)}\sim U(0,1)$, $\forall\:i$ and $\beta_i^{(0)}\sim U(0,1)$, $\forall\:i$. As for the UBCM, the identification $p_{ij}^\text{DBCM}\equiv\frac{e^{-\alpha_i^{(\infty)}-\beta_j^{(\infty)}}}{1+e^{-\alpha_i^{(\infty)}-\beta_j^{(\infty)}}}$, $\forall\:i\neq j$ allows the probability coefficients defining the DBCM to be numerically determined.\\

\paragraph{Reducing the dimensionality of the problem.} As for the UBCM, we can define a `reduced' version of the DBCM likelihood, accounting only for the \textit{distinct} (pairs of) values of the degrees. By defining $k^{out}\equiv k$ and $k^{in}\equiv h$, in order to simplify the formalism, the reduced DBCM recipe reads

\begin{eqnarray}\label{eq:dbcm_rd}
\mathscr{L}_\mathrm{DBCM}^\text{reduced}(\vec{\theta})&=&-\sum_k\sum_h n(k,h)[\alpha_{k,h}k(\mathbf{A}^*)+\beta_{k,h}h(\mathbf{A}^*)]\nonumber\\
&&-\sum_{k,h}\sum_{k',h'}n(k,h)[n(k',h')-\delta_{kk'}\delta_{hh'}]\cdot\nonumber\\
&&\cdot\ln\left[1+e^{-\alpha_{k,h}-\beta_{k',h'}}\right];
\end{eqnarray}
the implementation of the algorithms considered here must be modified in a way that is analogous to the one already described for the UBCM. In particular, the fixed-point recipe for the DBCM can be re-written by assigning to the nodes with the same out- and in-degrees $(k,h)$ the same pair of values $(\alpha,\beta)$, i.e. as

\begin{widetext}
\begin{eqnarray}
\label{eq:iterative_dcm}
\alpha^{(n)}_{k,h}&=&-\ln\left[\frac{k(\mathbf{A}^*)}{\sum_{k',h'}[n(k',h')-\delta_{kk'}\delta_{hh'}]\left(\frac{e^{-\beta_{k',h'}^{(n-1)}}}{1+e^{-\alpha_{k,h}^{(n-1)}-\beta_{k',h'}^{(n-1)}}}\right)}\right],\quad\forall\:k,h\\
\label{eq:iterative_dcm2}
\beta^{(n)}_{k,h}&=&-\ln\left[\frac{h(\mathbf{A}^*)}{\sum_{k',h'}[n(k',h')-\delta_{kk'}\delta_{hh'}]\left(\frac{e^{-\alpha_{k',h'}^{(n-1)}}}{1+e^{-\alpha_{k,h}^{(n-1)}-\beta_{k',h'}^{(n-1)}}}\right)}\right],\quad\forall\:k,h
\end{eqnarray}
\end{widetext}
where the sums, now, run over the \emph{distinct} values of the out- and in-degrees, $n(k,h)$ is the number of nodes whose out-degree is $k$ and whose in-degree is $h$ and, as usual, the last term at the denominator excludes the self-contribution (i.e. the probability that a node links to itself).\\

\begin{table*}[t!]
\begin{tabular}{|l|c|c|c|c||c|c||c|c||c|c|}
\hline
\multicolumn{1}{c}{} & \multicolumn{1}{c}{} & \multicolumn{1}{c}{} & \multicolumn{1}{c}{} & \multicolumn{1}{c}{} & \multicolumn{2}{c}{Newton} & \multicolumn{2}{c}{Quasi-Newton} & \multicolumn{2}{c}{Fixed-point}\\
\hline
& $N$ & $L$ & $c$ & $c_{r}$ & MADE & Time (s) & MADE & Time (s) & MADE & Time (s)\\
\hline
WTW 92 & $162$ & $5891$ & $\simeq 2.3 \cdot 10^{-1}$ & $\simeq 2.8 \cdot 10^{-1}$ & $\simeq 2.6 \cdot 10^{-9}$ & $\simeq 3$ & $\simeq 1.4 \cdot 10^{-10}$ & $\simeq 0.12$ & $\simeq 3.5 \cdot 10^{-2}$ & $\simeq 2.5$\\
\hline
WTW 93 & $162$ & $7384$ & $\simeq 2.8 \cdot 10^{-1}$ & $\simeq 3.5 \cdot 10^{-1}$ & $\simeq 6.7 \cdot 10^{-9}$ & $\simeq 0.5$ & $\simeq 5.3 \cdot 10^{-7}$ & $\simeq 0.16$ & $\simeq 3.4 \cdot 10^{-2}$ & $\simeq 3.1$\\
\hline
WTW 94 & $162$ & $9395$ & $\simeq 3.6 \cdot 10^{-1}$ & $\simeq 4.1 \cdot 10^{-1}$ & $\simeq 2.6 \cdot 10^{-9}$ & $\simeq 7.4$ & $\simeq 3.1 \cdot 10^{-10}$ & $\simeq 0.18$ & $\simeq 3.5 \cdot 10^{-2}$ & $\simeq 3.7$\\
\hline
WTW 95 & $162$ & $10947$ & $\simeq 4.2 \cdot 10^{-1}$ & $\simeq 4.3 \cdot 10^{-1}$ & $\simeq 4 \cdot 10^{-9}$ & $\simeq 16$ & $\simeq 4 \cdot 10^{-7}$ & $\simeq 0.21$ & $\simeq 3.3 \cdot 10^{-2}$ & $\simeq 3.9$\\
\hline
WTW 96 & $162$ & $11869$ & $\simeq 4.6 \cdot 10^{-1}$ & $\simeq 4.5 \cdot 10^{-1}$ & $\simeq 3.3 \cdot 10^{-9}$ & $\simeq 1.1$ & $\simeq 9.2 \cdot 10^{-7}$ & $\simeq 0.33$ & $\simeq 3.3 \cdot 10^{-2}$ & $\simeq 4.1$\\
\hline
WTW 97 & $162$ & $12840$ & $\simeq 4.9 \cdot 10^{-1}$ & $\simeq 4.5 \cdot 10^{-1}$ & $\simeq 2.4 \cdot 10^{-9}$ & $\simeq 13$ & $\simeq 2.4 \cdot 10^{-10}$ & $\simeq 0.16$ & $\simeq 3.3 \cdot 10^{-2}$ & $\simeq 4.2$\\
\hline
WTW 98 & $162$ & $13344$ & $\simeq 5.1 \cdot 10^{-1}$ & $\simeq 4.5 \cdot 10^{-1}$ & $\simeq 2.4 \cdot 10^{-9}$ & $\simeq 9.3$ & $\simeq 3.9 \cdot 10^{-10}$ & $\simeq 0.16$ & $\simeq 3.3 \cdot 10^{-2}$ & $\simeq 4.2$\\
\hline
WTW 99 & $162$ & $13810$ & $\simeq 5.3 \cdot 10^{-1}$ & $\simeq 4.5 \cdot 10^{-1}$ & $\simeq 2.5 \cdot 10^{-9}$ & $\simeq 13$ & $\simeq 9.4 \cdot 10^{-7}$ & $\simeq 0.3$ & $\simeq 3.3 \cdot 10^{-2}$ & $\simeq 4.1$\\
\hline
WTW 00 & $162$ & $14095$ & $\simeq 5.4 \cdot 10^{-1}$ & $\simeq 4.8 \cdot 10^{-1}$ & $\simeq 1.8 \cdot 10^{-9}$ & $\simeq 10$ & $\simeq 3.9 \cdot 10^{-10}$ & $\simeq 0.17$ & $\simeq 3.3 \cdot 10^{-2}$ & $\simeq 4.4$\\
\hline
WTW 01 & $162$ & $14521$ & $\simeq 5.6 \cdot 10^{-1}$ & $\simeq 4.7 \cdot 10^{-1}$ & $\simeq 6.8 \cdot 10^{-9}$ & $\simeq 10$ & $\simeq 4.2 \cdot 10^{-6}$ & $\simeq 0.22$ & $\simeq 3.4 \cdot 10^{-2}$ & $\simeq 4.3$\\
\hline
WTW 02 & $162$ & $13911$ & $\simeq 5.3 \cdot 10^{-1}$ & $\simeq 4.6 \cdot 10^{-1}$ & $\simeq 3.8 \cdot 10^{-9}$ & $\simeq 13$ & $\simeq 5.8 \cdot 10^{-10}$ & $\simeq 0.18$ & $\simeq 3.3 \cdot 10^{-2}$ & $\simeq 4.3$\\
\hline
\hline
$\text{BTC}_\text{week-1}^\text{CC}$ & 13576 & 20604 & $\simeq 1.1 \cdot 10^{-4}$ & $\simeq 1 \cdot10^{-2}$ & $\simeq 3\cdot 10^{-7}$ & $\simeq 1.3$ & $\simeq 1.5\cdot 10^{-4}$  & $\simeq 0.4$  & $\simeq 4 \cdot 10^{-6}$ & $\simeq 0.1$\\
\hline
$\text{BTC}_\text{week-1}$ & $20984$ & $25553$ & $\simeq 5.8 \cdot 10^{-5}$ & $\simeq 6.6 \cdot 10^{-3}$ & $\simeq 5.1 \cdot 10^{-4}$ & $\simeq 4$ & $\simeq 2.7 \cdot 10^{-7}$ & $\simeq 0.11$ & $\simeq 6\cdot 10^{-6}$ & $\simeq 0.05$\\
\hline
$\text{BTC}_\text{week-2}^\text{CC}$ & 297338 & 554643 & $\simeq 6 \cdot 10^{-6}$ & $\simeq 2 \cdot10^{-3}$ & $\simeq 9\cdot 10^{-9}$ & $\simeq 12$ & $\simeq 7.4 \cdot 10^{-5}$  & $\simeq 7$  & $\simeq 8 \cdot 10^{-6}$ & $\simeq 0.65$\\
\hline
$\text{BTC}_\text{week-2}$ & $338334$ & $571551$ & $\simeq 5 \cdot 10^{-6}$ & $\simeq 1.9 \cdot 10^{-3}$ & $\simeq 1.5 \cdot 10^{-4}$ & $\simeq 16$ & $\simeq 9.5 \cdot 10^{-4}$ & $\simeq 2.3$ & $\simeq 1.1 \cdot 10^{-5}$ & $\simeq 0.5$\\
\hline
\hline
Twitter & $436551$ & $1489857$ & $\simeq 7.8 \cdot 10^{-6}$ & $\simeq 4.6 \cdot 10^{-3}$ & $\simeq 9.5 \cdot 10^{-9}$ & $\simeq 6$ & $\simeq 4.9 \cdot 10^{-3}$ & $\simeq 40$ & $\simeq 1.3 \cdot 10^{-5}$ & $\simeq 7.6$\\
\hline
\end{tabular}
\caption{Performance of Newton's, quasi-Newton and the fixed-point algorithm to solve the reduced system of equations defining the DBCM, on a set of real-world BDNs (of which basic statistics as the total number of nodes, $N$, the total number of links, $L$, and the connectance, $c=L/N(N-1)$, are provided). For what concerns the World Trade Web, both Newton's and the quasi-Newton methods stop because the condition $||\nabla\mathscr{L}(\vec{\theta})||_2\leq 10^{-8}$ is satisfied; the fixed-point recipe, instead, always reaches the limit of 10000 steps. The fastest and most accurate method is systematically the quasi-Newton one. The picture changes when very large networks, as Bitcoin and Twitter, are considered: in these cases, the fastest and most accurate method is the fixed-point one. Only the results corresponding to the best choice of initial conditions are reported.}
\label{tab:directed}
\end{table*}

\paragraph{Performance testing.} As for the UBCM, the performance of the three algorithms in solving the reduced version of eqs. \eqref{sys:dbcm1} and \eqref{sys:dbcm2} has been tested on a bunch of real-world networks. The latter ones span economic, financial and social networks. In particular, we have considered the World Trade Web (WTW) during the decade 1992-2002 \cite{Squartini2011b}, a pair of snapshots of the Bitcoin User Network at the weekly time scale (the first day of those weeks being 13-02-12 and 27-04-15, respectively) \citep{Bovet2019} and of the corresponding largest weakly connected component (whose size is, respectively, $\simeq 65\%$ and $\simeq 90\%$ of the full network size) and a snapshot of the semantic network concerning the Twitter discussion about the Covid-19 pandemics (more precisely: it is the network of re-tweets of the (online) Italian debate about Covid-19, collected in the period $21^\text{st}$ February - $20^\text{th}$ April 2020) \citep{Caldarelli2020}. Before commenting on the results of our numerical exercises, let us, first, describe how the latter ones have been carried out.\\

The accuracy of each algorithm in reproducing the constraints defining the DBCM has been quantified via the maximum absolute error metrics that, in this case, reads

\begin{equation}
\text{MADE}=\max_i\{\left|k_i^*-\langle k_i\rangle\right|, \left|h_i^*-\langle h_i\rangle\right|\}_{i=1}^N
\end{equation}
and accounts for the presence of two different degrees per node. As for the UBCM, it is the infinite norm of the difference between the vector of the empirical values of the constraints and that of their expected values.

The three different `stop criteria' we have considered match the ones adopted for analysing the undirected case and consist in a condition on the Euclidean norm of the gradient of the likelihood function, i.e. $||\nabla\mathscr{L}(\vec{\theta})||_2\leq 10^{-8}$, in a condition on the Euclidean norm of the vector of differences between the values of the parameters at subsequent iterations, i.e. $||\Delta\vec{\theta}||_2\leq 10^{-8}$, and in a condition on the maximum number of iterations: after 10000 steps, any of the three algorithms stops.\\

The results about the performance of our three algorithms are reported in \textcolor{black}{Table \ref{tab:directed}}. Overall, all recipes perform very satisfactorily, being accurate, fast and scalable; however, while Newton's and the quasi-Newton methods stop because the condition on the norm of the likelihood is satisfied, the fixed-point recipe is always found to satisfy the limit condition on the number of steps (i.e. it runs for 10000 steps and, then, stops).

Let us start by commenting the results concerning the WTW. For what concerns accuracy, the largest maximum error per method spans an interval, across all configurations, that amounts at $\text{MADE}_\text{Newton}^\text{reduced}\simeq 10^{-9}$, $10^{-10}\lesssim\text{MADE}_\text{Quasi-Newton}^\text{reduced}\lesssim 10^{-6}$ and $\text{MADE}_\text{fixed-point}^\text{reduced}\simeq 10^{-2}$. By looking at each specific network, it is evident that the most accurate method is systematically the quasi-Newton one.

For what concerns speed, the amount of time required by each method to achieve convergence spans an interval (across all configurations) that is $0.5\leq T_\text{Newton}^\text{reduced}\leq 13$, $0.21\leq T_\text{Quasi-Newton}^\text{reduced}\leq 0.33$ and $2.5\leq T_\text{fixed-point}^\text{reduced}\leq 4.4$ (time is measured in seconds). The fastest method is the quasi-Newton one, followed by Newton's method and the fixed-point recipe. The latter is the slowest method since it always reaches the limit of 10000 steps while the other two competing ones stop after few iterations. Appreciable differences in the speed of convergence of any method, caused by the choice of a particular set of initial conditions, are not observed.

The observations above hold true when the WTW is considered. The picture changes when very large networks, as Bitcoin and Twitter, are considered. First, let us notice that Bitcoin and Twitter `behave' as the undirected version of Internet considered to solve the UBCM, i.e. they are very redundant, hosting many nodes with the same out- and in-degrees \textcolor{black}{(in fact, the coefficient of reduction, $c_r$, is, now, defined as the number of different `out-degree - in-degree' pairs divided by twice the number of nodes).} To provide a specific example, out of the original $676688$ equations defining the DBCM for one of the two Bitcoin snapshots considered here, only $\simeq339$ equations survive the reduction; by converse, the WTW can be reduced to a much smaller extent (to be more specific, out of the original $324$ equations defining the DBCM for the WTW in 1997, only $\simeq291$ equations survive the reduction). Interestingly, a good proxy of the reducibility of the directed configurations considered here is provided by their connectance (i.e. the denser the network, the less reducible it is).

On the one hand, this feature, common to the very large networks considered here, is what guarantees their resolution in a reasonable amount of time; on the other one, it seems not to be enough to let Newton's and the quasi-Newton method be as fast as in the undirected case. For our binary, directed networks, in fact, the fastest (and, for some configurations, the most accurate) method becomes the fixed-point one. In order to understand this result we need to consider that both Newton's and the quasi-Newton method require (some proxy of) the Hessian matrix of the DBCM to update the value of the parameters: since the order of the latter is $O(N^2)$ for Newton's method and $O(N)$ for the quasi-Newton one, its calculation can be (very) time demanding - beside requiring a lot of memory for the step-wise update of the corresponding Hessian matrix. However, while this is compensated by a larger accuracy in the case of Newton's method, this is no longer true when the quasi-Newton recipe is considered - the reason maybe lying in the poorer approximation provided by the diagonal of the Hessian matrix in case of systems like these.\\

As a last comment, we would like to stress that, as in the undirected case, the generic coefficient $p_{ij}^\text{DBCM}$ represents a proper probability, in turn implying that eq. \eqref{dcm} also provides us with a recipe to sample the canonical ensemble of BDNs, under the DBCM. Notice that the factorization of the graph probability $P_\text{DBCM}(\mathbf{A}|\vec{\theta})$ greatly simplifies the entire procedure, allowing a single graph to be sampled by implementing the Bernoulli trial

\begin{equation}
a_{ij}=
\begin{cases}
0 \: & 1-p_{ij}^\text{DBCM} \\
1 \: & p_{ij}^\text{DBCM}
\end{cases}
\end{equation}
for each (directed) pair of nodes, in either a sequential or a parallel fashion. The sampling process, whose computational complexity amounts at $O(N^2)$, can be repeated to generate as many configurations as desired. The pseudo-code for explicitly sampling the DBCM ensemble is summed up by Algorithm 2.

\begin{algorithm}[H]
\begin{algorithmic}[1]
\For{$m=1\dots |E|$}
\State $\mathbf{A}=\mathbf{0};$
\For{$i=1\dots N$}
\For{$j=1\dots N$ \textrm{and} $j\neq i$}
\If{$\text{RandomUniform}[0,1]\leq p_{ij}^\text{DBCM}$}
\State $a_{ij}=1;$
\Else
\State $a_{ij}=0;$
\EndIf
\EndFor
\EndFor
\State $\text{Ensemble}[m]=\mathbf{A};$
\EndFor
\caption{Sampling the DBCM ensemble}
\label{code:directed}
\end{algorithmic}
\end{algorithm}

\subsection{BiCM: bipartite binary undirected graphs\\with given degree sequences}\label{sec:BiCM}

So far, we have considered monopartite networks. However, the algorithm we have described for solving the DBCM can be adapted, with little effort, to solve a null model designed for bipartite, binary, undirected networks (BiBUNs), i.e. the so-called \emph{Bipartite Configuration Model} (BiCM) \cite{Saracco2015}. These networks are defined by two distinct layers (say, $\top$ and $\bot$) and obey the rule that links can exist only between (and not within) layers: for this reason, they can be compactly described via a biadjacency matrix $\mathbf{B}\equiv\{b_{i\alpha}\}_{i,\alpha}$ whose generic entry $b_{i\alpha}$ is 1 if node $i$ belonging to layer $\bot$ is linked to node $\alpha$ belonging to layer $\top$ and 0 otherwise. The constraints defining the BiCM are represented by the degree sequences $\{k_i\}_{i\in\bot}$ and $\{d_\alpha\}_{\alpha\in\top}$ where $k_i=\sum_{\alpha\in\top} b_{i\alpha}$ counts the neighbors of node $i$ (belonging to layer $\top$) and $d_\alpha=\sum_{i\in\bot}b_{i\alpha}$ counts the neighbors of node $\alpha$ (belonging to layer $\bot$).

Analogously to the DBCM case,

\begin{equation}\label{eq:bicm}
P(\mathbf{B}|\vec{\gamma},\vec{\beta})=\prod_{i\in\bot}\prod_{\alpha\in\top}p_{i\alpha}^{b_{i\alpha}}(1-p_{i\alpha})^{1-b_{i\alpha}}
\end{equation}
where $p_{i\alpha}=p_{i\alpha}^\text{BiCM}\equiv\frac{e^{-\gamma_i-\beta_{\alpha}}}{1+e^{-\gamma_i-\beta_{\alpha}}}$. The canonical ensemble of BiBUNs includes all networks with, say, $N$ nodes on one layer, $M$ nodes on the other layer and a number of links (connecting nodes of different layers) ranging from zero to the maximum value $N\cdot M$.

\begin{table*}[t!]
\begin{tabular}{|l|c|c|c|c||c|c||c|c||c|c|}
\hline
\multicolumn{5}{c}{} & \multicolumn{2}{c}{Newton} & \multicolumn{2}{c}{Quasi-Newton} & \multicolumn{2}{c}{Fixed-point}\\
\hline
& $N+M$ & $L$ & $c$ & $c_r$ & MADE & Time (s) & MADE & Time (s) & MADE & Time (s)\\
\hline
 WTW 95 & 1277 & 18947 & $\simeq 1.1 \cdot 10^{-1}$ & $\simeq 1.3 \cdot 10^{-1}$ & $\simeq 1.1 \cdot 10^{-13}$ & $\simeq 0.0022$ & $\simeq 3.0 \cdot 10^{-6}$ & $\simeq 0.012$ & $\simeq 7.7 \cdot 10^{-6}$ & $\simeq 0.012$ \\
 \hline 
 WTW 96 & 1277 & 19934 & $\simeq 1.2 \cdot 10^{-1}$ & $\simeq 1.3 \cdot 10^{-1}$ & $\simeq 2.3 \cdot 10^{-13}$ & $\simeq 0.0023$ & $\simeq 1.5 \cdot 10^{-6}$ & $\simeq 0.014$ & $\simeq 1.1 \cdot 10^{-4}$ & $\simeq 0.023$ \\
 \hline 
 WTW 97 & 1277 & 20222 & $\simeq 1.2 \cdot 10^{-1}$ & $\simeq 1.3 \cdot 10^{-1}$ & $\simeq 1.7 \cdot 10^{-13}$ & $\simeq 0.0022$ & $\simeq 3.5 \cdot 10^{-6}$ & $\simeq 0.02$ & $\simeq 2.4 \cdot 10^{-4}$ & $\simeq 0.013$ \\
 \hline 
 WTW 98 & 1277 & 20614 & $\simeq 1.2 \cdot 10^{-1}$ & $\simeq 1.4 \cdot 10^{-1}$ & $\simeq 2.8 \cdot 10^{-13}$ & $\simeq 0.0024$ & $\simeq 1.2 \cdot 10^{-6}$ & $\simeq 0.015$ & $\simeq 1.8 \cdot 10^{-4}$ & $\simeq 0.018$ \\
 \hline 
 WTW 99 & 1277 & 20949 & $\simeq 1.3 \cdot 10^{-1}$ & $\simeq 1.4 \cdot 10^{-1}$ & $\simeq 2.3 \cdot 10^{-13}$ & $\simeq 0.0024$ & $\simeq 2.8 \cdot 10^{-5}$ & $\simeq 0.012$ & $\simeq 2.1 \cdot 10^{-4}$ & $\simeq 0.019$ \\
 \hline 
 WTW 00 & 1277 & 21257 & $\simeq 1.3 \cdot 10^{-1}$ & $\simeq 1.4 \cdot 10^{-1}$ & $\simeq 2.3 \cdot 10^{-13}$ & $\simeq 0.0025$ & $\simeq 1.3 \cdot 10^{-6}$ & $\simeq 0.016$ & $\simeq 2.8 \cdot 10^{-5}$ & $\simeq 0.018$ \\
 \hline 
 WTW 01 & 1277 & 21326 & $\simeq 1.3 \cdot 10^{-1}$ & $\simeq 1.3 \cdot 10^{-1}$ & $\simeq 1.7 \cdot 10^{-13}$ & $\simeq 0.0023$ & $\simeq 3.4 \cdot 10^{-5}$ & $\simeq 0.015$ & $\simeq 2.5 \cdot 10^{-5}$ & $\simeq 0.015$ \\
 \hline 
 WTW 02 & 1277 & 21333 & $\simeq 1.3 \cdot 10^{-1}$ & $\simeq 1.4 \cdot 10^{-1}$ & $\simeq 1.7 \cdot 10^{-13}$ & $\simeq 0.0024$ & $\simeq 4.1 \cdot 10^{-6}$ & $\simeq 0.018$ & $\simeq 2.1 \cdot 10^{-4}$ & $\simeq 0.016$ \\
 \hline 
 WTW 03 & 1277 & 21330 & $\simeq 1.3 \cdot 10^{-1}$ & $\simeq 1.3 \cdot 10^{-1}$ & $\simeq 2.8 \cdot 10^{-13}$ & $\simeq 0.0023$ & $\simeq 1.1 \cdot 10^{-6}$ & $\simeq 0.015$ & $\simeq 4.3 \cdot 10^{-5}$ & $\simeq 0.014$ \\
 \hline 
 WTW 04 & 1277 & 21479 & $\simeq 1.3 \cdot 10^{-1}$ & $\simeq 1.3 \cdot 10^{-1}$ & $\simeq 1.7 \cdot 10^{-13}$ & $\simeq 0.0024$ & $\simeq 2.2 \cdot 10^{-7}$ & $\simeq 0.018$ & $\simeq 2.1 \cdot 10^{-4}$ & $\simeq 0.019$ \\
 \hline 
 WTW 05 & 1278 & 21841 & $\simeq 1.3 \cdot 10^{-1}$ & $\simeq 1.4 \cdot 10^{-1}$ & $\simeq 2.3 \cdot 10^{-13}$ & $\simeq 0.0024$ & $\simeq 2.2 \cdot 10^{-6}$ & $\simeq 0.013$ & $\simeq 2.3 \cdot 10^{-4}$ & $\simeq 0.027$ \\
 \hline 
 WTW 06 & 1279 & 21945 & $\simeq 1.3 \cdot 10^{-1}$ & $\simeq 1.3 \cdot 10^{-1}$ & $\simeq 2.3 \cdot 10^{-13}$ & $\simeq 0.0023$ & $\simeq 1.3 \cdot 10^{-5}$ & $\simeq 0.016$ & $\simeq 2.2 \cdot 10^{-4}$ & $\simeq 0.012$ \\
 \hline 
 WTW 07 & 1279 & 22036 & $\simeq 1.3 \cdot 10^{-1}$ & $\simeq 1.4 \cdot 10^{-1}$ & $\simeq 2.3 \cdot 10^{-13}$ & $\simeq 0.0024$ & $\simeq 2.0 \cdot 10^{-6}$ & $\simeq 0.017$ & $\simeq 2.1 \cdot 10^{-4}$ & $\simeq 0.023$ \\
 \hline 
 WTW 08 & 1279 & 21889 & $\simeq 1.3 \cdot 10^{-1}$ & $\simeq 1.3 \cdot 10^{-1}$ & $\simeq 1.1 \cdot 10^{-13}$ & $\simeq 0.0023$ & $\simeq 1.5 \cdot 10^{-5}$ & $\simeq 0.017$ & $\simeq 2.5 \cdot 10^{-4}$ & $\simeq 0.024$ \\
 \hline 
 WTW 09 & 1279 & 21621 & $\simeq 1.3 \cdot 10^{-1}$ & $\simeq 1.3 \cdot 10^{-1}$ & $\simeq 2.3 \cdot 10^{-13}$ & $\simeq 0.0025$ & $\simeq 2.1 \cdot 10^{-6}$ & $\simeq 0.021$ & $\simeq 2.4 \cdot 10^{-4}$ & $\simeq 0.018$ \\
 \hline 
 WTW 10 & 1279 & 21010 & $\simeq 1.3 \cdot 10^{-1}$ & $\simeq 1.3 \cdot 10^{-1}$ & $\simeq 2.3 \cdot 10^{-13}$ & $\simeq 0.0022$ & $\simeq 1.6 \cdot 10^{-6}$ & $\simeq 0.015$ & $\simeq 2.6 \cdot 10^{-4}$ & $\simeq 0.022$ \\
\hline
\end{tabular}
\caption{Performance of Newton's, quasi-Newton and the fixed-point algorithm to solve the reduced system of equations defining the BiCM, on a set of real-world BiUNs (of which basic statistics as the total number of nodes, $N$, the total number of links, $L$, and the connectance, $c=L/(N\cdot M)$, are provided). All algorithms stop because the condition $||\nabla\mathscr{L}(\vec{\theta})||_2\leq 10^{-10}$ is satisfied. For what concerns both accuracy and speed, the best performing method is Newton's one, followed by the quasi-Newton and the fixed-point recipes. Only the results corresponding to the best choice of initial conditions are reported.}
\label{tab:bipartite}
\end{table*}

The BiCM likelihood function reads

\begin{eqnarray}\label{bicm_likelihood}
\mathscr{L}_\text{BiCM}(\vec{\gamma},\vec{\beta})=&-&\sum\limits_{i\in\bot} \gamma_ik_i(\mathbf{B^*})-\sum\limits_{\alpha\in\top}\beta_\alpha d_\alpha(\mathbf{B}^*)\nonumber\\
&-& \sum\limits_{i\in\bot}\sum\limits_{\alpha\in\top}\ln\left[1+e^{-\gamma_i-\beta_\alpha}\right]
\end{eqnarray}
whose first-order optimality conditions read

\begin{eqnarray}  
\nabla_{\gamma_i}\mathscr{L}_\text{BiCM}&=&-k_i(\mathbf{B}^*)+\sum_{\alpha \in \top} \dfrac{e^{-\gamma_i-\beta_\alpha}}{1+e^{-\gamma_i-\beta_\alpha}},\quad i\in\bot\nonumber\\
\nabla_{\beta_\alpha}\mathscr{L}_\text{BiCM}&=&-d_\alpha(\mathbf{B}^*)+\sum_{i \in \bot} \dfrac{e^{-\gamma_i-\beta_\alpha}}{1+e^{-\gamma_i-\beta_\alpha}},\quad\alpha\in\top.\nonumber\\
\end{eqnarray}

\paragraph{Resolution of the BiCM.} As for the DBCM case, Newton's and the quasi-Newton methods can be implemented by adapting the recipe defined in eq. \eqref{eq:gensol} to the bipartite case (see Appendix A for the definition of the BiCM Hessian).

As for the DBCM, the fixed-point recipe for the BiCM can be re-written in the usual iterative fashion as follows

\begin{eqnarray}
\gamma_i^{(n)}=-\ln\left[\frac{k_i(\mathbf{B}^*)}{\sum_{\alpha \in \top}\left(\frac{e^{-\beta_\alpha^{(n-1)}}}{1+e^{-\gamma_i^{(n-1)}-\beta_\alpha^{(n-1)}}}\right)}\right],\quad\forall\:i\in\bot\nonumber\\
\beta_\alpha^{(n)}=-\ln\left[\frac{d_\alpha(\mathbf{B}^*)}{\sum_{i \in \bot}\left(\frac{e^{-\gamma_i^{(n-1)}}}{1+e^{-\gamma_i^{(n-1)}-\beta_\alpha^{(n-1)}}}\right)}\right],\quad\forall\:\alpha\in\top\nonumber\\
\end{eqnarray}
and the initialization is similar as well: in fact, we can employ the value of the solution of the BiCM in the sparse case, i.e. $\gamma_i^{(0)}=-\ln\left[ \frac{k_i(\mathbf{B}^*)}{\sqrt{L}}\right]$, $\forall\:i\in\bot$ and $\beta_\alpha^{(0)}=-\ln\left[\frac{d_\alpha(\mathbf{B}^*)}{\sqrt{L}}\right]$, $\forall\:\alpha\in\top$ (only this set of initial conditions has been employed to analyse the bipartite case).\\

\paragraph{Reducing the dimensionality of the problem.} Exactly as for the DBCM case, the presence of nodes with the same degree, on the same layer, leads to the appearance of identical equations in the system above; hence, the computation of the solutions can be sped up by writing

\begin{eqnarray}\label{sys:bicm}
\gamma_k^{(n)}=-\ln\left[\frac{k(\mathbf{B}^*)}{\sum_{d}g(d)\left(\frac{e^{-\beta_{k,d}^{(n-1)}}}{1+e^{-\gamma_{k,d}^{(n-1)}-\beta_{k,d}^{(n-1)}}}\right)}\right],\quad\forall\:k\nonumber\\
\beta_d^{(n)}=-\ln\left[\frac{d(\mathbf{B}^*)}{\sum_{k}f(k)\left(\frac{e^{-\gamma_{k,d}^{(n-1)}}}{1+e^{-\gamma_{k,d}^{(n-1)}-\beta_{k,d}^{(n-1)}}}\right)}\right],\quad\forall\:d\nonumber\\
\end{eqnarray}
where $f(k)$ is the number of nodes, belonging to layer $\bot$, whose degree is $k$ and $g(d)$ is the number of nodes, belonging to layer $\top$, whose degree is $d$.\\

\paragraph{Performance testing.} The performance of the three algorithms in solving eqs. \eqref{sys:bicm} has been tested on 16 snapshot of the bipartite, binary, undirected version of the WTW, gathering the country-product export relationships across the years 1995-2010 \cite{Saracco2015}. Before commenting on the results of our numerical exercises, let us, first, describe how the latter ones have been carried out.\\

The accuracy of each algorithm in reproducing the constraints defining the BiCM has been quantified via the maximum absolute error metrics that, now, reads

\begin{eqnarray}
\text{MADE}=\max&\{&\left|k_1^*-\langle
k_1\rangle\right|\dots\left|k_N^*-\langle k_N\rangle\right|,\nonumber\\
&&\left|d_1^*-\langle d_1\rangle\right|\dots\left|d_M^*-\langle d_M\rangle\right|\}
\end{eqnarray}
to account for the degrees of nodes on both layers.

The three different `stop criteria' match the ones adopted for analysing the UBCM and the DBCM and consist in a condition on the Euclidean norm of the gradient of the likelihood function, i.e. $||\nabla\mathscr{L}(\vec{\theta})||_2\leq 10^{-10}$, in a condition on the Euclidean norm of the vector of differences between the values of the parameters at subsequent iterations, i.e. $||\Delta\vec{\theta}||_2\leq 10^{-10}$, and a condition on the maximum number of iterations: after 1000 steps, any of the three algorithms stops.\\

The results about the performance of our three algorithms are reported in \textcolor{black}{Table \ref{tab:bipartite}}. Overall, all recipes are accurate, fast and scalable; all methods stop because the condition on the norm of the likelihood is satisfied.

For what concerns accuracy, the largest maximum error per method spans an interval (across all configurations) that amounts at $\text{MADE}_\text{Newton}^\text{reduced}\simeq 10^{-13}$, $10^{-7}\lesssim\text{MADE}_\text{Quasi-Newton}^\text{reduced}\lesssim 10^{-5}$ and $10^{-5}\lesssim\text{MADE}_\text{fixed-point}^\text{reduced}\lesssim 10^{-4}$. By looking at each specific network, it is evident that the most accurate method is systematically Newton's one.

For what concerns speed, the amount of time required by each method to achieve convergence spans an interval (across all configurations) that is $T_\text{Newton}^\text{reduced}\simeq 0.0023$ (on average), $T_\text{Quasi-Newton}^\text{reduced}\simeq 0.016$ (on average) and $T_\text{fixed-point}^\text{reduced}\simeq 0.018$ (on average) (time is measured in seconds). The fastest method is Newton's one and is followed by the quasi-Newton and the fixed-point recipes. The gain in terms of speed due to the reducibility \textcolor{black}{(now quantified by the ratio between the number of different pairs of degrees divided by the total number of nodes)} of the system of equations defining the BiCM is also evident: while solving the original problem would have required handling a system of $\simeq 10^3$ equations, the reduced one is defined by just $\simeq 10^2$ distinct equations. Overall, a solution is always found within thousandths or hundredths of seconds.\\

As for the DBCM case, the ensemble of BiBUNs can be sampled by implementing a Bernoulli trial $b_{i\alpha}\sim\text{Ber}[p_{i\alpha}]$ for any two nodes (belonging to different layers) in either a sequential or a parallel fashion. The computational complexity of the sampling process amounts at $O(N\cdot M)$ and can be repeated to generate as many configurations as desired. The pseudo-code for explicitly sampling the BiCM ensemble is summed up by Algorithm 3.

\begin{algorithm}[H]
\begin{algorithmic}[1]
\For{$m=1\dots |E|$}
\State $\mathbf{B}=\mathbf{0};$
\For{$i=1\dots N$}
\For{$\alpha=1\dots M$}
\If{$\text{RandomUniform}[0,1]\leq p_{i\alpha}^\text{BiCM}$}
\State $b_{i\alpha}=1;$
\Else
\State $b_{i\alpha}=0;$
\EndIf
\EndFor
\EndFor
\State $\text{Ensemble}[m]=\mathbf{B};$
\EndFor
\caption{Sampling the BiCM ensemble}
\label{code:bipartite}
\end{algorithmic}
\end{algorithm}

\subsection{UECM: weighted undirected graphs\\with given strengths and degrees\label{sec:UECM}}

So far, we have considered purely binary models. Let us now focus on a class of `mixed' null models for weighted networks, defined by constraining both binary and weighted quantities. Purely weighted models such as the \emph{Undirected} and the \emph{Directed Weighted Configuration Model} have not been considered since, as it has been proven elsewhere \cite{Mastrandrea2014a}, they perform quite poorly when employed to reconstruct networks. Let us start by the simplest model, i.e. the one constraining the degrees and the strengths in an undirected fashion. While $k_i(\mathbf{A})=\sum_{j(\neq i)=1}^Na_{ij}$ counts the number of neighbors of node $i$, $s_i(\mathbf{W})=\sum_{j(\neq i)=1}^Nw_{ij}$ defines the weighted equivalent of the degree of node $i$, i.e. its strength. For consistency, the binary adjacency matrix can be defined via the Heaviside step function, $\Theta[.]$, as $\mathbf{A}\equiv\Theta[\mathbf{W}]$ a position indicating that $a_{ij}=1$ if $w_{ij}>0$, $\forall\:i<j$ and zero otherwise. This particular model is known as \emph{Undirected Enhanced Configuration Model} (UECM) \cite{Garlaschelli2009,Mastrandrea2014a,Mastrandrea2014b} and its Hamiltonian reads

\begin{equation}
\mathcal{H}_\text{UECM}(\mathbf{W},\vec{\alpha},\vec{\beta})=\sum_{i=1}^N[\alpha_ik_i(\mathbf{A})+\beta_is_i(\mathbf{W})];
\end{equation}
it induces a probability distribution which is halfway between a Bernoulli and a geometric one \cite{Garlaschelli2009}, i.e.

\begin{equation}
Q_\text{UECM}(\mathbf{W}|\vec{\alpha},\vec{\beta})=\prod_{i=1}^N\prod_{\substack{j=1\\(j<i)}}^Nq_{ij}(w_{ij})
\end{equation}
with

\begin{equation}\label{eq:uecm}
q_{ij}(w)=
\begin{cases}
1-p_{ij}^\text{UECM}, \: & w=0\\
p_{ij}^\text{UECM}(e^{-\beta_i-\beta_j})^{w-1}(1-e^{-\beta_i-\beta_j}), \: & w>0
\end{cases}\\
\end{equation}
for any two nodes $i$ and $j$ such that $i<j$ and $p_{ij}^\text{UECM}=\frac{e^{-\alpha_i-\alpha_j-\beta_i-\beta_j}}{1-e^{-\beta_i-\beta_j}+e^{-\alpha_i-\alpha_j-\beta_i-\beta_j}}$. Notice that the functional form above is obtained upon requiring that the weights only assume (non-negative) integer values (i.e. $w_{ij}\in[0,+\infty)$, $\forall\:i<j$): hence, the canonical ensemble is now constituted by the weighted configurations with $N$ nodes and a number of (undirected) links ranging between zero and the maximum value $\binom{N}{2}$.

\begin{table*}[t!]
\begin{tabular}{|l|c|c|c||c|c|c|c||c|c|c|c|}
\hline
\multicolumn{4}{c}{} & \multicolumn{4}{c}{Newton} & \multicolumn{4}{c}{Quasi-Newton}\\
\hline
& $N$ & $L$ & $c$ & MRDE & MASE & MRSE & Time (s) & MRDE & MASE & MRSE & Time (s)\\
\hline
 WTW 90 & 169 & 7991 & $ \simeq 0.3 $ & $\simeq 2.6 \cdot 10^{-10}$ & $\simeq 5 \cdot 10^{-6} $ & $\simeq 2 \cdot 10^{-10}$ & $\simeq 0.4$ & $\simeq 7.8 \cdot 10^{-4}$ & $\simeq 2 \cdot 10^{-1} $ & $\simeq 3 \cdot 10^{-4}$ & $\simeq 25$\\
 \hline 
 WTW 91 & 184 & 8712 & $ \simeq 0.3 $ & $\simeq 2.1 \cdot 10^{-10}$ & $\simeq 1 \cdot 10^{-10}$ & $\simeq 1.2 \cdot 10^{-10}$ & $\simeq 0.5$ & $\simeq 9.2 \cdot 10^{-5}$ & $\simeq 7 \cdot 10^{-2}$ & $\simeq 6.6 \cdot 10^{-5}$ & $\simeq 28$\\
 \hline 
 WTW 92 & 185 & 8928 & $ \simeq 0.3 $ & $\simeq 1.2 \cdot 10^{-10}$ & $\simeq 7 \cdot 10^{-7}$ & $\simeq 1.3 \cdot 10^{-10}$ & $\simeq 0.5$ & $\simeq 7 \cdot 10^{-2}$ & $\simeq 2 \cdot 10^{-4}$ & $\simeq 1.3 \cdot 10^{-4}$ & $\simeq 28$\\
 \hline 
 WTW 93 & 187 & 9220 & $ \simeq 0.3 $ & $\simeq 3.4 \cdot 10^{-6}$ & $\simeq 2 \cdot 10^{-10}$ & $\simeq 2.2 \cdot 10^{-10}$ & $\simeq 0.5$ & $\simeq 1.4 \cdot 10^{-4}$ & $\simeq 1 \cdot 10^{-1}$ & $\simeq 7.6 \cdot 10^{-5}$ & $\simeq 28$\\
 \hline 
 WTW 94 & 187 & 9437 & $ \simeq 0.3 $ & $\simeq 1.8 \cdot 10^{-10}$ & $\simeq 2 \cdot 10^{-10}$ & $\simeq 1.8 \cdot 10^{-10}$ & $\simeq 0.7$ & $\simeq 7 \cdot 10^{-5}$ & $\simeq 2 \cdot 10^{-1}$ & $\simeq 1.5 \cdot 10^{-4}$ & $\simeq 28$\\
 \hline 
 WTW 95 & 187 & 9578 & $ \simeq 0.3 $ & $\simeq 2.8 \cdot 10^{-10}$ & $\simeq 3 \cdot 10^{-6}$ & $\simeq 2.9 \cdot 10^{-10}$ & $\simeq 0.6$ & $\simeq 2.5 \cdot 10^{-4}$ & $\simeq 3 \cdot 10^{-1}$ & $\simeq 6.7 \cdot 10^{-5}$ & $\simeq 28$\\
 \hline 
 WTW 96 & 187 & 10002 & $ \simeq 0.3 $ & $\simeq 1.4 \cdot 10^{-5}$ & $\simeq 1 \cdot 10^{-10}$ & $\simeq 1.1 \cdot 10^{-10}$ & $\simeq 0.7$ & $\simeq 1.7 \cdot 10^{-5}$ & $\simeq 2 \cdot 10^{-1}$ & $\simeq 3 \cdot 10^{-5}$ & $\simeq 28$\\
 \hline 
 WTW 97 & 187 & 10251 &$ \simeq 0.3 $ & $\simeq 1.1 \cdot 10^{-5}$ & $\simeq 4 \cdot 10^{-10}$ & $\simeq 6.7 \cdot 10^{-10}$ & $\simeq 0.7 $ & $\simeq 4.4 \cdot 10^{-5}$ & $\simeq 8 \cdot 10^{-2}$ & $\simeq 1.6 \cdot 10^{-4}$ & $\simeq 28$\\
 \hline 
 WTW 98 & 187 & 10254 & $ \simeq 0.3 $ & $\simeq 1 \cdot 10^{-5}$ & $\simeq 3 \cdot 10^{-10}$ & $\simeq 4 \cdot 10^{-10}$ & $\simeq 0.6$ & $\simeq 1.7 \cdot 10^{-4}$ & $\simeq 8 \cdot 10^{-2}$ & $\simeq 5.3 \cdot 10^{-5}$ & $\simeq 28$\\
 \hline 
 WTW 99 & 187 & 10252 & $ \simeq 0.3 $ & $\simeq 4.7 \cdot 10^{-10}$ & $\simeq 1 \cdot 10^{-10}$ & $\simeq 8 \cdot 10^{-10}$ & $\simeq 0.7$ & $\simeq 1.6 \cdot 10^{-4}$ & $\simeq 7 \cdot 10^{-2}$ & $\simeq 6.2 \cdot 10^{-5}$ & $\simeq 28$\\
 \hline 
 WTW 00 & 187 & 10252 & $ \simeq 0.3 $ & $\simeq 5 \cdot 10^{-10}$ & $\simeq 2 \cdot 10^{-10}$ & $\simeq 2.4 \cdot 10^{-10}$ & $\simeq 0.7$ & $\simeq 1.5 \cdot 10^{-4}$ & $\simeq 9 \cdot 10^{-2}$ & $\simeq 5.4 \cdot 10^{-5}$ & $\simeq 29$\\
\hline
\end{tabular}
\caption{Performance of Newton's and the quasi-Newton method to solve the reduced system of equations defining the UECM, on a set of real-world WUNs (of which basic statistics as the total number of nodes, $N$, the total number of links, $L$, and the connectance, $c=2L/N(N-1)$, are provided). While Newton's method stops because the condition $||\nabla\mathscr{L}(\vec{\theta})||_2\leq 10^{-8}$ is satisfied, the quasi-Newton one always reaches the limit of 10000 steps. The results on accuracy and speed clearly indicate that Newton's method outperforms the quasi-Newton one. Only the results corresponding to the best choice of initial conditions are reported. The results of the fixed-point recipe are not shown.}
\label{tab:uecm}
\end{table*}

The argument of the problem \eqref{eq:nlp} for the specific network $\mathbf{W}^*$ now becomes

\begin{eqnarray}
\mathscr{L}_\mathrm{UECM}(\vec{\alpha},\vec{\beta})=&-&\sum_{i=1}^N[\alpha_ik_i(\mathbf{A}^*)+\beta_is_i(\mathbf{W}^*)]\nonumber\\
&-&\sum_{i=1}^N\sum_{\substack{j=1\\(j<i)}}^N\ln\left[1+e^{-\alpha_i-\alpha_j}\left(\frac{e^{-\beta_i-\beta_j}}{1-e^{-\beta_i-\beta_j}}\right)\right]\nonumber\\
\end{eqnarray}
whose first-order optimality conditions read

\begin{eqnarray}
\nabla_{\alpha_i}\mathscr{L}_\mathrm{UECM}&=&-k_i(\mathbf{A}^*)+\sum_{j(\neq i)=1}^Np_{ij}^\text{UECM}\nonumber\\
&=&-k_i(\mathbf{A}^*)+\langle k_i\rangle=0,\quad i=1\dots N,\nonumber\\
\nabla_{\beta_i}\mathscr{L}_\mathrm{UECM}&=&-s_i(\mathbf{W}^*)+\sum_{j(\neq i)=1}^N\frac{p_{ij}^\text{UECM}}{1-e^{-\beta_i-\beta_j}}\nonumber\\
&=&-s_i(\mathbf{W}^*)+\langle s_i\rangle=0,\quad i=1\dots N.\nonumber\\
\end{eqnarray}

\paragraph{Resolution of the UECM.} Newton's and the quasi-Newton methods can be easily implemented via the recipe defined in eq. \eqref{eq:gensol} (see Appendix A for the definition of the UECM Hessian).

As for the purely binary models, the fixed-point recipe for solving the UECM first-order optimality conditions transforms the following set of consistency equations

\begin{eqnarray}
\alpha_i&=&-\ln\left[\frac{k_i(\mathbf{A}^*)}{\sum_{\substack{j=1\\(j\neq i)}}^N\left(\frac{e^{-\alpha_j-\beta_i-\beta_j}}{1-e^{-\beta_i-\beta_j}+e^{-\alpha_i-\alpha_j-\beta_i-\beta_j}}\right)}\right],\nonumber\\
\beta_i&=&-\ln\left[\frac{s_i(\mathbf{W}^*)}{\sum_{\substack{j=1\\(j\neq i)}}^N\left(\frac{e^{-\alpha_i-\alpha_j-\beta_j}}{(1-e^{-\beta_i-\beta_j})(1-e^{-\beta_i-\beta_j}+e^{-\alpha_i-\alpha_j-\beta_i-\beta_j})}\right)}\right]\nonumber\\
\end{eqnarray}
(with $i=1\dots N$) into the usual iterative fashion, by considering the parameters at the left hand side and at the right hand side, respectively at the $n$-th and at the $(n-1)$-th iteration. It is important to remark that a reduced version of the iterative recipe above can indeed be written, by assigning the same pair of values $(\alpha,\beta)$ to the nodes with the same pair of values $(k,s)$: however, the larger heterogeneity of the strengths causes this event to happen more rarely than for purely binary models such as the UBCM and the DBCM.

As for the purely binary cases, three different sets of initial conditions have been considered, whose definition follows from the simplest conceivable generalization of the purely binary cases. In particular, the first set of values reads $\alpha_i^{(0)}=-\ln\left[\frac{k_i(\mathbf{A}^*)}{\sqrt{2L}}\right]$, $i=1\dots N$ and $\beta_i^{(0)}=-\ln\left[\frac{s_i(\mathbf{W}^*)}{\sqrt{2W}}\right]$, $i=1\dots N$; the second set is a variant of the first, reading $\alpha_i^{(0)}=-\ln\left[\frac{k_i(\mathbf{A}^*)}{\sqrt{N}}\right]$, $i=1\dots N$ and $\beta_i^{(0)}=-\ln\left[\frac{s_i(\mathbf{W}^*)}{\sqrt{N}}\right]$, $i=1\dots N$; the third recipe, instead, prescribes to randomly draw the value of each parameter from the uniform distribution defined on the unit interval, i.e. $\alpha_i^{(0)}\sim U(0,1)$, $\forall\:i$ and $\beta_i^{(0)}\sim U(0,1)$, $\forall\:i$.

\paragraph{Performance testing.} The performance of the three algorithms to solve the system of equations defining the UECM has been tested on a bunch of real-world networks. In particular, we have considered the WTW during the decade 1990-2000 \cite{Gleditsch2002}. Since the weights defining the configurations of the WTW are real numbers, we have rounded them to the nearest integer value, before running the UECM. Before commenting on the results of our numerical exercises, let us, first, describe how the latter ones have been carried out.\\

The accuracy of each algorithm in reproducing the constraints defining the UECM has been now quantified via the \emph{maximum relative error} metrics, defined, in a perfectly general fashion, as $\max_i\left\{\frac{|C_i^*-\langle C_i\rangle|}{C_i}\right\}_{i=1}^N$ (where $C_i^*$ is the empirical value of the $i$-th constraint, $C_i$). In the UECM case, we can define two variants of the aforementioned error, i.e.

\begin{eqnarray}
\text{MRDE}&=&\max_i\left\{\frac{|k_i^*-\langle k_i\rangle|}{k_i}\right\}_{i=1}^N\label{eq:errsa}\\
\text{MRSE}&=&\max_i\left\{\frac{|s_i^*-\langle s_i\rangle|}{s_i}\right\}_{i=1}^N\label{eq:errsb}
\end{eqnarray}
(the acronyms standing for Maximum Relative Degree Error and Maximum Relative Strength Error). The reason driving this choice lies in the evidence that, in absolute terms, strengths are affected by a larger numerical error than degrees: this, however, doesn't necessarily mean that a given algorithm performs poorly, as the magnitude of an error must be always compared with the numerical value of the quantity it refers to - whence the choice of considering relative scores.

The three different `stop criteria' we have considered for each algorithm match the ones adopted for analysing the binary cases, consisting in a condition on the Euclidean norm of the gradient of the likelihood function, i.e. $||\nabla\mathscr{L}(\vec{\theta})||_2\leq 10^{-8}$, and in a condition on the Euclidean norm of the vector of differences between the values of the parameters at subsequent iterations, i.e. $||\Delta\vec{\theta}||_2\leq 10^{-8}$. The third condition concerns the maximum number of iterations: after 10000 steps, any of the three algorithms stops.\\

The results about the performance of our three algorithms are reported in \textcolor{black}{Table \ref{tab:uecm}}. Overall, two out of three algorithms (i.e. Newton's and the quasi-Newton methods) perform very satisfactorily, being accurate, fast and scalable; the third one (i.e. the fixed-point recipe), instead, performs very poorly. Moreover, while Newton's method stops because the condition on the norm of the likelihood is satisfied, both the quasi-Newton and the fixed-point algorithms are always found to satisfy the limit condition on the number of steps (i.e. they run for 10000 steps and, then, stop).

For what concerns accuracy, the largest maximum error made by Newton's method (across all configurations) amounts at $10^{-10}\lesssim\text{MRDE}_\text{Newton}\lesssim 10^{-5}$ and $\text{MRSE}_\text{Newton}\gtrsim 10^{-10}$; on the other hand, the largest maximum error made by the quasi-Newton method (across all configurations) amounts at $10^{-5}\lesssim\text{MRDE}_\text{Quasi-Newton}\lesssim 10^{-1}$ and $10^{-5}\leq\text{MRSE}_\text{Quasi-Newton}\leq 10^{-4}$. For what concerns speed, Newton's method employs \textit{tenths of seconds} to achieve convergence on each configuration while the quasi-Newton one always requires \textit{tens of seconds} (specifically, almost thirty seconds for each considered configuration). The results above indicate that the fastest and two most accurate method is systematically Newton's one, suggesting that the `complexity' of the model is such that the information encoded into the Hessian matrix cannot be ignored without consequences on the quality of the solution. The fixed-point algorithm, instead, stops after seconds but is affected by errors whose order of systematically magnitude amounts at $\text{MRDE}_\text{fixed-point}\simeq 10^2$ and $1\lesssim\text{MRSE}_\text{fixed-point}\lesssim10^2$.

We also explicitly notice that the MADE basically coincides with the MRDE for all considered configurations, meaning that the largest error, made by the algorithms considered here to solve the UECM, affects the nodes with the lowest degree (i.e. equal to one). On the other hand, strengths are affected by a larger absolute error (i.e. the MASE, defined as $\text{MASE}=\max_i\{\left|s_i^*-\langle s_i \rangle\right|\}_{i=1}^N$) than the degrees: if we calculate the MRSE, however, we realize that the largest errors affect very large strengths - hence being perfectly acceptable. For example, let us consider the WTW in 1993: the MASE amounts at 0.1 but, as the MRSE reveals, it affects a strength of the order of $10^3$.

Lastly, differences in the speed of convergence of the two methods discussed in this section, caused by the choice of a particular set of initial conditions, are observable: the `uniform' prescription outperforms the other ones.\\

Finally, let us comment on the algorithm to sample the UECM ensemble and that can be compactly achieved by implementing a two-step procedure. Let us look back at the formal expression for the pair-specific probability distribution characterizing the UECM: it induces coefficients reading

\begin{equation}
\begin{cases}
1-p_{ij}^\text{UECM},\:& w=0\\
p_{ij}^\text{UECM}(1-e^{-\beta_i-\beta_j}),\:& w=1\\
p_{ij}^\text{UECM}(e^{-\beta_i-\beta_j})(1-e^{-\beta_i-\beta_j}),\:& w=2\\
p_{ij}^\text{UECM}(e^{-\beta_i-\beta_j})^2(1-e^{-\beta_i-\beta_j}),\:& w=3\\
\hspace{2.53cm}\vdots
\end{cases}
\end{equation}
in turn suggesting that, for a specific pair of vertices $i,j$ (with $i<j$), the appearance of the first link is ruled by a Bernoulli distribution with probability $p_{ij}^\text{UECM}$ while the remaining $(w-1)$ ones can be drawn from a geometric distribution whose parameter reads $e^{-\beta_i-\beta_j}$; in other words, the weight $(w-1)$ is drawn \emph{conditionally} on the presence of a connection between the two considered nodes. The computational complexity of the sampling process is, again, $O(N^2)$. The pseudo-code for explicitly sampling the DBCM ensemble is summed up by Algorithm 4. Notice that the way our sampling procedure is written requires the support of the geometric distribution to coincide with the positive integers.

\begin{algorithm}[H]
\begin{algorithmic}[1]
\For{$m=1\dots |E|$}
\State $\mathbf{W}=\mathbf{0};$
\For{$i=1\dots N$}
\For{$j=1\dots N$ \textrm{and} $j<i$}
\If{$\text{RandomUniform}[0,1]\leq p_{ij}^\text{UECM}$}
\State $w_{ij}=w_{ji}=\text{RandomGeometric}[e^{-\beta_i-\beta_j}];$
\Else
\State $w_{ij}=w_{ji}=0;$
\EndIf
\EndFor
\EndFor
\State $\text{Ensemble}[m]=\mathbf{W};$
\EndFor
\caption{Sampling the UECM ensemble}
\label{code:uecm}
\end{algorithmic}
\end{algorithm}

\subsection{DECM: weighted directed graphs\\with given strengths and degrees\label{sec:DECM}}

Let us now extend the `mixed' model introduced in the previous section to the case of directed networks. Constraints are, now, represented by four sequences of values, i.e. $\{k_i^{out}\}_{i=1}^N$, $\{k_i^{in}\}_{i=1}^N$, $\{s_i^{out}\}_{i=1}^N$, $\{s_i^{in}\}_{i=1}^N$ where the generic out-degree and in-degree are, respectively, defined as $k_i^{out}(\mathbf{A})=\sum_{j(\neq i)=1}^Na_{ij}$ and $k_i^{in}(\mathbf{A})=\sum_{j(\neq i)=1}^Na_{ji}$ and analogously for the generic out-strength and in-strength, reading $s_i^{out}(\mathbf{W})=\sum_{j(\neq i)=1}^Nw_{ij}$ and $s_i^{in}(\mathbf{W})=\sum_{j(\neq i)=1}^Nw_{ji}$. Consistency requires that $\mathbf{A}\equiv\Theta[\mathbf{W}]$ as for the UECM case. This model is known as \emph{Directed Enhanced Configuration Model} (DECM) and its Hamiltonian reads

\begin{eqnarray}
\mathcal{H}_\text{DECM}(\mathbf{W},\vec{\alpha},\vec{\beta},\vec{\gamma},\vec{\delta})&=&\sum_{i=1}^N[\alpha_ik_i^{out}(\mathbf{A})+\beta_ik_i^{in}(\mathbf{A})\nonumber\\
&&+\gamma_is_i^{out}(\mathbf{W})+\delta_is_i^{in}(\mathbf{W})]\nonumber\\
\end{eqnarray}
in turn, inducing the directed counterpart of the UECM distribution, i.e.

\begin{equation}
Q_\text{DECM}(\mathbf{W}|\vec{\alpha},\vec{\beta},\vec{\gamma},\vec{\delta})=\prod_{i=1}^N\prod_{\substack{j=1\\(j\neq i)}}^Nq_{ij}(w_{ij})
\end{equation}
with

\begin{equation}
q_{ij}(w)=
\begin{cases}
1-p_{ij}^\text{DECM} \: & w=0\\
p_{ij}^\text{DECM}(e^{-\gamma_i-\delta_j})^{w-1}(1-e^{-\gamma_i-\delta_j)} \: & w>0
\end{cases}\\
\end{equation}
for any two nodes $i$ and $j$ such that $i\neq j$ and $p_{ij}^\text{DECM}=\frac{e^{-\alpha_i-\beta_j-\gamma_i-\delta_j}}{1-e^{-\gamma_i-\delta_j}+e^{-\alpha_i-\beta_j-\gamma_i-\delta_j}}$. As for the undirected case, weights are required to assume only (non-negative) integer values (i.e. $w_{ij}\in[0,+\infty)$, $\forall\:i\neq j$): hence, the canonical ensemble is constituted by the weighted configurations with $N$ nodes and a number of (directed) links ranging between zero and the maximum value $N(N-1)$.

The argument of the problem \eqref{eq:nlp} for the specific network $\mathbf{W}^*$ becomes

\begin{eqnarray}
\mathscr{L}_\mathrm{DECM}(\vec{\alpha},\vec{\beta},\vec{\gamma},\vec{\delta})=&-&\sum_{i=1}^N[\alpha_ik_i^{out}(\mathbf{A}^*)+\beta_ik_i^{in}(\mathbf{A}^*)\nonumber\\
&+&\gamma_is_i^{out}(\mathbf{W}^*)+\delta_is_i^{in}(\mathbf{W}^*)]\nonumber\\
&-&\sum_{i=1}^N\sum_{\substack{j=1\\(j\neq i)}}^N\ln z_{ij}
\end{eqnarray}
where $z_{ij}=\left[1+e^{-\alpha_i-\beta_j}\left(\frac{e^{-\gamma_i-\delta_j}}{1-e^{-\gamma_i-\delta_j}}\right)\right]$, $\:\forall\:i\neq j$ and whose first-order optimality conditions read

\begin{eqnarray}
\nabla_{\alpha_i}\mathscr{L}_\mathrm{DECM}&=&-k_i^{out}(\mathbf{A}^*)+\sum_{j(\neq i)=1}^Np_{ij}^\text{DECM}\nonumber\\
&=&-k_i^{out}(\mathbf{A}^*)+\langle k_i^{out}\rangle=0,\quad i=1\dots N,\nonumber\\
\nabla_{\beta_i}\mathscr{L}_\mathrm{DECM}&=&-k_i^{in}(\mathbf{A}^*)+\sum_{j(\neq i)=1}^Np_{ji}^\text{DECM}\nonumber\\
&=&-k_i^{in}(\mathbf{A}^*)+\langle k_i^{in}\rangle=0,\quad i=1\dots N,\nonumber\\
\nabla_{\gamma_i}\mathscr{L}_\mathrm{DECM}&=&-s_i^{out}(\mathbf{W}^*)+\sum_{j(\neq i)=1}^N\frac{p_{ij}^\text{DECM}}{1-e^{-\gamma_i-\delta_j}}\nonumber\\
&=&-s_i^{out}(\mathbf{W}^*)+\langle s_i^{out}\rangle=0,\quad i=1\dots N,\nonumber\\
\nabla_{\delta_i}\mathscr{L}_\mathrm{DECM}&=&-s_i^{in}(\mathbf{W}^*)+\sum_{j(\neq i)=1}^N\frac{p_{ji}^\text{DECM}}{1-e^{-\gamma_j-\delta_i}}\nonumber\\
&=&-s_i^{in}(\mathbf{W}^*)+\langle s_i^{in}\rangle=0,\quad i=1\dots N.\nonumber\\
\end{eqnarray}

\paragraph{Resolution of the DECM.} Newton's and the quasi-Newton methods can be easily implemented via the recipe defined in eq. \eqref{eq:gensol} (see Appendix A for the definition of the DECM Hessian).

\begin{table*}[t!]
\begin{tabular}{|l|c|c|c||c|c|c|c||c|c|c|c|}
\hline
\multicolumn{4}{c}{} & \multicolumn{4}{c}{Newton} & \multicolumn{4}{c}{Quasi-Newton}\\
\hline
& $N$ & $L$ & $c$ & MRDE & MASE & MRSE & Time (s) & MRDE & MASE & MRSE & Time (s)\\
\hline
 e-MID 00 & 196 & 10618 & $\simeq 0.28 $ &
 $\simeq 1.4 \cdot 10^{-10}$ & $\simeq 5 \cdot 10^{-9}$ & $\simeq 1.5 \cdot 10^{-10}$ & $\simeq 0.5 $ &
 $\simeq 1.7 \cdot 10^{-7}$ & $\simeq 3 \cdot 10^{-1}$ & $\simeq 5.4 \cdot 10^{-7}$ & $\simeq 0.1$\\
 \hline 
 e-MID 01 & 185 & 8951 & $\simeq 0.26 $ &
 $\simeq 1.4 \cdot 10^{-11}$ & $\simeq 6 \cdot 10^{-9}$ & $\simeq 2 \cdot 10^{-10}$ & $\simeq 0.4 $ &
 $\simeq 1.4 \cdot 10^{-7}$ & $\simeq 10$ & $\simeq 7 \cdot 10^{-5}$ & $\simeq 0.2$\\
 \hline 
 e-MID 02 & 177 & 7252 & $\simeq 0.23 $ &
 $\simeq 1.4 \cdot 10^{-15}$ & $\simeq \cdot 10^{-4}$ & $\simeq 1 \cdot 10^{-5}$ & $\simeq 0.5 $ &
 $\simeq 9.5 \cdot 10^{-8}$ & $\simeq 6 \cdot 10^{-1}$ & $\simeq 7.4 \cdot 10^{-6}$ & $\simeq 0.1$\\
 \hline 
 e-MID 03 & 179 & 6814 & $\simeq 0.21 $ &
 $\simeq 1.6 \cdot 10^{-10}$ & $\simeq 2 \cdot 10^{-5}$ & $\simeq 4.4 \cdot 10^{-10}$ & $\simeq 0.9 $ &
 $\simeq 9.6 \cdot 10^{-8}$ & $\simeq 50$ & $\simeq 1.1 \cdot 10^{-3}$ & $\simeq 0.2 $\\
 \hline 
 e-MID 04 & 180 & 6136 & $\simeq 0.19 $ &
 $\simeq 6.5 \cdot 10^{-13}$ & $\simeq 9 \cdot 10^{-7}$ & $\simeq 3.4 \cdot 10^{-12}$ & $\simeq 0.9 $ &
 $\simeq 1 \cdot 10^{-7}$ & $\simeq 700$ & $\simeq 4.2 \cdot 10^{-3}$ & $\simeq 0.2$\\
 \hline 
 e-MID 05 & 176 & 6203 & $\simeq 0.2 $ &
 $\simeq 3 \cdot 10^{-12}$ & $\simeq 1 \cdot 10^{-5}$ & $\simeq 9.4 \cdot 10^{-11}$ & $\simeq 1.2 $ &
 $\simeq 4.8 \cdot 10^{-8}$ & $\simeq 300$ & $\simeq 2.4 \cdot 10^{-3}$ & $\simeq 0.3$\\
 \hline 
 e-MID 06 & 177 & 6132 & $\simeq 0.19 $ &
 $\simeq 1.5 \cdot 10^{-14}$ & $\simeq 2 \cdot 10^{-7}$ & $\simeq 1.8 \cdot 10^{-11}$ & $\simeq 0.9 $ &
 $\simeq 5 \cdot 10^{-8}$ & $\simeq 60$ & $\simeq 2.5 \cdot 10^{-3}$ & $\simeq 0.2$\\
 \hline 
 e-MID 07 & 178 & 6330 & $\simeq 0.2 $ &
 $\simeq 8.4 \cdot 10^{-15}$ & $\simeq 1 \cdot 10^{-4}$ & $\simeq 2.5 \cdot 10^{-6}$ & $\simeq 0.7 $ &
 $\simeq 1.8 \cdot 10^{-8}$ & $\simeq 3$ & $\simeq 7.6 \cdot 10^{-5}$ & $\simeq 0.1 $\\
 \hline 
 e-MID 08 & 173 & 4767 & $\simeq 0.16 $ &
 $\simeq 2.3 \cdot 10^{-9}$ & $\simeq 2 \cdot 10^{-9}$ & $\simeq 7.7 \cdot 10^{-10}$ & $\simeq 0.6 $ &
 $\simeq 1.8 \cdot 10^{-8}$ & $\simeq 20$ & $\simeq 1.2 \cdot 10^{-3}$ & $\simeq 0.3$\\
 \hline 
 e-MID 09 & 156 & 2961 & $\simeq 0.12 $ &
 $\simeq 2.6 \cdot 10^{-15}$ & $\simeq 1 \cdot 10^{-7}$ & $\simeq 4.9 \cdot 10^{-12}$ & $\simeq 0.6 $ &
 $\simeq 1.7 \cdot 10^{-8}$ & $\simeq 1$ & $\simeq 9 \cdot 10^{-5}$ & $\simeq 0.1 $\\
 \hline 
 e-MID 10 & 135 & 2743 & $\simeq 0.15 $ &
 $\simeq 6.3 \cdot 10^{-8}$ & $\simeq 3 \cdot 10^{-6}$ & $\simeq 5.9 \cdot 10^{-10}$ & $\simeq 0.7 $ &
 $\simeq 1.6 \cdot 10^{-8}$ & $\simeq 4$ & $\simeq 5.2 \cdot 10^{-5}$ & $\simeq 0.1$\\
\hline
\end{tabular}
\caption{Performance of Newton's and the quasi-Newton method to solve the reduced system of equations defining the DECM, on a set of real-world WDNs (of which basic statistics as the total number of nodes, $N$, the total number of links, $L$, and the connectance, $c=L/N(N-1)$, are provided). While Newton's method stops because the condition $||\nabla\mathscr{L}(\vec{\theta})||_2\leq 10^{-8}$ is satisfied, the quasi-Newton one always reaches the limit of 10000 steps. The results on accuracy and speed clearly indicate that Newton's method outperforms the quasi-Newton one. Only the results corresponding to the best choice of initial conditions are reported. The results of the fixed-point recipe are not shown.}
\label{tab:decm}
\end{table*}

As for the UECM, the fixed-point recipe for solving the DECM first-order optimality conditions transforms the following set of consistency equations

\begin{eqnarray}
\alpha_i&=&-\ln\left[\frac{k_i^{out}(\mathbf{A}^*)}{\sum_{\substack{j=1\\(j\neq i)}}^N\left(\frac{e^{-\beta_j-\gamma_i-\delta_j}}{1-e^{-\gamma_i-\delta_j}+e^{-\alpha_i-\beta_j-\gamma_i-\delta_j}}\right)}\right],\nonumber\\
\beta_i&=&-\ln\left[\frac{k_i^{in}(\mathbf{A}^*)}{\sum_{\substack{j=1\\(j\neq i)}}^N\left(\frac{e^{-\alpha_j-\gamma_j-\delta_i}}{1-e^{-\gamma_j-\delta_i}+e^{-\alpha_j-\beta_i-\gamma_j-\delta_i}}\right)}\right],\nonumber\\
\gamma_i&=&-\ln\left[\frac{s_i^{out}(\mathbf{W}^*)}{\sum_{\substack{j=1\\(j\neq i)}}^N\left(\frac{e^{-\alpha_i-\beta_j-\delta_j}}{(1-e^{-\gamma_i-\delta_j})(1-e^{-\gamma_i-\delta_j}+e^{-\alpha_i-\beta_j-\gamma_i-\delta_j})}\right)}\right],\nonumber\\
\delta_i&=&-\ln\left[\frac{s_i^{in}(\mathbf{W}^*)}{\sum_{\substack{j=1\\(j\neq i)}}^N\left(\frac{e^{-\alpha_j-\beta_i-\gamma_j}}{(1-e^{-\gamma_j-\delta_i})(1-e^{-\gamma_j-\delta_i}+e^{-\alpha_j-\beta_i-\gamma_j-\delta_i}}\right)}\right].\nonumber\\
\end{eqnarray}
(with $i=1\dots N$) into the usual iterative fashion, by considering the parameters at the left hand side and at the right hand side, respectively at the $n$-th and at the $(n-1)$-th iteration. The reduced version of such a recipe would assign the same set of values $(\alpha,\beta,\gamma,\delta)$ to the nodes for which the quantities $(k^{out},k^{in},s^{out},s^{in})$ have the same value: however, the larger heterogeneity of the strengths causes the DECM to be much less reducible than the purely binary models we have considered in the present contribution.

The three different sets of initial conditions that have been considered generalize the UECM ones: in particular, the first set of values reads $\alpha_i^{(0)}=-\ln\left[\frac{k_i^{out}(\mathbf{A}^*)}{\sqrt{L}}\right]$, $i=1\dots N$, $\beta_i^{(0)}=-\ln\left[\frac{k_i^{in}(\mathbf{A}^*)}{\sqrt{L}}\right]$, $i=1\dots N$, $\gamma_i^{(0)}=-\ln\left[\frac{s_i^{out}(\mathbf{W}^*)}{\sqrt{W}}\right]$, $i=1\dots N$ and $\delta_i^{(0)}=-\ln\left[\frac{s_i^{in}(\mathbf{W}^*)}{\sqrt{W}}\right]$, $i=1\dots N$; the second set of initial conditions can be obtained by simply replacing $L$ with $N$; the third recipe, as usual, prescribes to randomly draw the value of each parameter from the uniform distribution defined on the unit interval.\\

\paragraph{Performance testing.} The performance of the three algorithms to solve the system of equations defining the DECM has been tested on a bunch of real-world networks. In particular, we have considered the Electronic Italian Interbank Market (e-MID) during the decade 2000-2010 \cite{Iori2006}. Since e-MID weights are real numbers, we have rounded them to the nearest integer value, before running the DECM. Before commenting on the results of our numerical exercises, let us, first, describe how the latter ones have been carried out.\\

The accuracy of each algorithm in reproducing the constraints defining the DECM has been quantified via the \emph{maximum relative error} metrics, now reading

\begin{eqnarray}
\text{MRDE}&=&\max_i\left\{\frac{|k_i^*-\langle k_i\rangle|}{k_i}, \frac{|h_i^*-\langle h_i\rangle| }{h_i}\right\}_{i=1}^N\label{eq:errs2a}\\
\text{MRSE}&=&\max_i\left\{\frac{|s_i^*-\langle s_i\rangle|}{s_i}, \frac{|t_i^*-\langle t_i\rangle|}{t_i}\right\}_{i=1}^N\label{eq:errs2b}
\end{eqnarray}
(the acronyms standing for for Maximum Relative Degree Error and Maximum Relative Strength Error) where we have defined $k^{out}\equiv k$, $k^{in}\equiv h$, $s^{out}\equiv s$ and $s^{in}\equiv t$ in order to simplify the formalism.

The three different `stop criteria' we have adopted are the same ones we have considered for both the binary and the undirected, `mixed' model, i.e. the condition on the Euclidean norm of the gradient of the likelihood function, i.e. $||\nabla\mathscr{L}(\vec{\theta})||_2\leq 10^{-8}$), the condition on the Euclidean norm of the vector of differences between the values of the parameters at subsequent iterations (i.e. $||\Delta\vec{\theta}||_2\leq 10^{-8}$) and the condition on the maximum number of iterations (i.e. after 10000 steps, any of the three algorithms stops).\\

The results about the performance of our three algorithms are reported in \textcolor{black}{Table \ref{tab:decm}}. Overall, Newton's method performs very satisfactorily, being accurate, fast and scalable; the quasi-Newton method is accurate as well although (in some cases, much) slower. The fixed-point recipe, instead, performs very poorly, as for the undirected case. Moreover, while Newton's method stops because the condition on the norm of the likelihood is satisfied, both the quasi-Newton and the fixed-point algorithms are always found to satisfy the limit condition on the number of steps (i.e. they run for 10000 steps and, then, stop).

For what concerns accuracy, the largest maximum error made by Newton's method (across all configurations) amounts at $10^{-14}\lesssim\text{MRDE}_\text{Newton}\lesssim 10^{-7}$ and $10^{-12}\lesssim\text{MRSE}_\text{Newton}\lesssim 10^{-5}$; on the other hand, the largest maximum error made by the quasi-Newton method (across all configurations) amounts at $10^{-8}\lesssim\text{MRDE}_\text{Quasi-Newton}\lesssim 10^{-7}$ and $10^{-4}\lesssim\text{MRSE}_\text{Quasi-Newton}\lesssim 10^{-3}$. For what concerns speed, Newton's method employs tens of seconds to achieve convergence on each configuration; the time required by the quasi-Newton method is of the same order of magnitude, although it is systematically larger than the time required by Newton's one. Overall, these results indicate that the fastest and two most accurate method is Newton's one. As in the undirected case, the fixed-point algorithm, instead, stops after seconds but is affected by errors whose order of systematically magnitude amounts at $10\lesssim\text{MRDE}_\text{fixed-point}\lesssim10^2$ and $1\lesssim\text{MRSE}_\text{fixed-point}\lesssim10^2$.

As for the UECM, the MADE basically coincides with the MRDE, for all considered configurations, while strengths are affected by a larger absolute error than the degrees: still, upon calculating the MRSE, we realize that the largest errors affect very large strengths - hence being perfectly acceptable.

Lastly, differences in the speed of convergence of the two methods discussed in this section, caused by the choice of a particular set of initial conditions, are observable: the `uniform' prescription outperforms the other ones.\\

Finally, let us comment on the algorithm to sample the DECM ensemble: as for the UECM, it can be compactly achieved by implementing the directed counterpart of the two-step procedure described above. Given a specific pair of vertices $i,j$ (with $i\neq j$), the first link can be drawn by sampling a Bernoulli distribution with probability $p^\text{DECM}_{ij}$ while the remaining $(w-1)$ ones can be drawn from a geometric distribution whose parameter reads $e^{-\gamma_i-\delta_j}$. The computational complexity of the sampling process is, again, $O(N^2)$ and the pseudo-code for explicitly sampling the DECM ensemble is summed up by Algorithm 5. Notice that the way our sampling procedure is written requires the support of the geometric distribution to coincide with the positive integers.

\begin{algorithm}[H]
\begin{algorithmic}[1]
\For{$m=1\dots |E|$}
\State $\mathbf{W}=\mathbf{0};$
\For{$i=1\dots N$}
\For{$j=1\dots N$ \textrm{and} $j\neq i$}
\If{$\text{RandomUniform}[0,1]\leq p_{i\alpha}^\text{DECM}$}
\State $w_{ij}=\text{RandomGeometric}[e^{-\gamma_i-\delta_j}];$
\Else
\State $w_{ij}=0;$
\EndIf
\EndFor
\EndFor
\State $\text{Ensemble}[m]=\mathbf{W};$
\EndFor
\caption{Sampling the DECM ensemble}
\label{code:decm}
\end{algorithmic}
\end{algorithm}

\subsection{Two-step models for\\undirected and directed networks}

The need of considering network models defined in a two-step fashion arises from a number of considerations. First, the amount of information concerning binary and weighted quantities is often asymmetric: as it has been pointed out in \cite{Parisi2020}, information concerning a given network structure ranges from the knowledge of just a single, aggregated piece of information (e.g. the link density) to that of entire subgraphs. Indeed, models exist that take as input any binary, either probabilistic or deterministic, network model (i.e. any $P(\mathbf{A})$) while placing link weights optimally, \emph{conditionally} on the input configurations \cite{Gabrielli2019,Parisi2020}.

Second, recipes like the UECM and the DECM are, generally speaking, difficult to solve; as we have already observed, only Newton's method performs in a satisfactory way, both for what concerns accuracy and speed: hence, easier-to-solve recipes are welcome.

In what follows, we will consider the conditional reconstruction method (hereby, CReM) induced by the Hamiltonian

\begin{eqnarray}
\mathcal{H}_\text{CReM}(\mathbf{W},\vec{\theta})&=&\sum_{i=1}^N\theta_is_i(\mathbf{A});
\end{eqnarray}
in case of undirected networks, it induces a conditional probability distribution reading

\begin{equation}
\mathbf{Q}(\mathbf{W}|\mathbf{A})=\prod_{i=1}^N\prod_{\substack{j=1\\(j<i)}}^Nq_{ij}(w_{ij}|a_{ij})
\end{equation}
where, for consistency, $q_{ij}(w_{ij}=0|a_{ij}=0)=1$ and $q_{ij}(w_{ij}=0|a_{ij}=1)=0$. The meaning of these relationships is the following: given any two nodes $i$ and $j$, the absence of a link, i.e. $a_{ij}=0$, admits the only possibility $w_{ij}=0$; on the other hand, the presence of a link, i.e. $a_{ij}=1$, rules out the possibility that a null weight among the same vertices is observed.

\begin{table*}[t!]
\centering
\begin{tabular}{|l||c|c|c|c||c|c|c|c|
|c|c|c|c|}
\hline
\multicolumn{1}{c}{} & \multicolumn{4}{c}{Newton} & \multicolumn{4}{c}{Quasi-Newton} & \multicolumn{4}{c}{Fixed-point} \\
\hline
& MRDE & MASE & MRSE & Time (s) & MRDE & MASE & MRSE & Time (s) & MRDE & MASE & MRSE & Time (s)\\
\hline
$\text{BLN}_1$ & $\simeq 4 \cdot 10^{-8}$ & $\simeq 10^{-9}$ & $\simeq 10^{-4}$ & $\simeq 0.08$ & $\simeq 6 \cdot 10^{-8}$ & $\simeq 10^{-4}$ & $\simeq 10^{-6}$ & $\simeq 5$ & $\simeq 2 \cdot 10^{-9}$ & $\simeq 10^{-2}$ & $\simeq 10^{-1}$ & $\simeq 0.01$ \\
\hline
$\text{BLN}_2$ & $\simeq 2 \cdot 10^{-8}$ & $\simeq 10^{-8}$ & $\simeq 10^{-5}$ & $\simeq 3.2$ & $\simeq 9 \cdot 10^{-8}$ & $\simeq 10^{-4}$ & $\simeq 10^{-1}$ & $\simeq 100$ & $\simeq 1 \cdot 10^{-9}$ & $\simeq 10^{-2}$ & $\simeq 10^{-1}$ & $\simeq 0.73$ \\
\hline
$\text{BLN}_3$ & $\simeq 1 \cdot 10^{-8}$ & $\simeq 10^{-9}$ & $\simeq 10^{-4}$ & $\simeq 14$ & $\simeq 1 \cdot 10^{-7}$ & $\simeq 10^{-4}$ & $\simeq 10^{-2}$ & $\simeq 388 $ & $\simeq 1 \cdot 10^{-9}$ & $\simeq 10^{-7}$ & $\simeq 10^{-6}$ & $\simeq 11$ \\
\hline
$\text{BLN}_4$ & $\simeq 2 \cdot 10^{-8}$ & $\simeq 10^{-9}$ & $\simeq 10^{-4}$ & $\simeq 71$ & $\simeq 5 \cdot 10^{-8}$ & $\simeq 10^{-2}$ & $\simeq 3$ & $\simeq 1538$ & $\simeq 9 \cdot 10^{-10}$ & $\simeq 20$ & $\simeq 6 \cdot 10^{-1}$ & $\simeq 1.3$ \\
\hline
$\text{BLN}_5$ & $\simeq 2 \cdot 10^{-9}$ & $\simeq 10^{-9}$ & $\simeq 10^{-4}$ & $\simeq 200 $ & $\simeq 4 \cdot 10^{-7}$ & $\simeq 10^{-1}$ & $\simeq 6$ & $\simeq 3633$ & $\simeq 6 \cdot 10^{-10}$ & $\simeq 10^{-7}$ & $\simeq 10^{-8}$ & $\simeq 5.7 $\\
\hline
$\text{BLN}_6$ & $\simeq 2 \cdot 10^{-9}$ & $\simeq 10^{-9}$ & $\simeq 10^{-4}$ & $\simeq 382 $ & $\simeq 3 \cdot 10^{-8}$ & $\simeq 10^{-2}$ & $\simeq 3$ & $\simeq 5980 $ & $\simeq 6 \cdot 10^{-10}$ & $\simeq 10^{-3}$ & $\simeq 10^{-3}$ & $\simeq 550$ \\
\hline
$\text{BLN}_7$ & $\simeq 5 \cdot 10^{-8}$ & $\simeq 10^{-9}$ & $\simeq 10^{-4}$ & $\simeq 648$ & $\simeq 4 \cdot 10^{-7}$ & $\simeq 10^{-2}$ & $\simeq 2$ & $\simeq 10177$ & $\simeq 5 \cdot 10^{-10}$ & $\simeq 10^{-2}$ & $\simeq 10^{-1}$ & $\simeq 36$ \\
\hline
$\text{BLN}_8$ & $\simeq 5 \cdot 10^{-12}$ & $\simeq 10^{-10}$ & $\simeq 10^{-6}$ & $\simeq 1188$ & $\simeq 3 \cdot 10^{-7}$ & $\simeq 10^{-4}$ & $\simeq 10^{-1}$ & $\simeq 15888 $ & $\simeq 5 \cdot 10^{-10}$ & $ \simeq 10^{-2}$ & $\simeq 10^{-1}$ & $\simeq 70$\\
\hline
\end{tabular}
\caption{Performance of Newton's, quasi-Newton and the fixed-point algorithm to solve the system of equations defining the (undirected version of the) CReM, on a set of real-world WUNs. While Newton's and the fixed-point method stop because the condition $||\nabla\mathscr{L}(\vec{\theta})||_2\leq 10^{-8}$ is satisfied, the quasi-Newton one often reaches the limit of 10000 steps. The results on accuracy and speed indicate that Newton's and the fixed-point method compete, outperforming the quasi-Newton one. Only the results corresponding to the best choice of initial conditions are reported.}
\label{tab:crem}
\end{table*}

In general, the functional form of $q_{ij}(w_{ij}|a_{ij}=1)$ depends on the domain of the weights. In all cases considered in \cite{Gabrielli2019,Parisi2020}, weights are assumed to be continuous; since the continuous distribution that maximizes Shannon entropy while constrained to reproduce first-order moments is the exponential one, the following functional form

\begin{equation}
q_{ij}(w_{ij}|a_{ij}=1)=
\begin{cases}
(\theta_i+\theta_j)e^{-(\theta_i+\theta_j)w}\: & w>0 \\
\:0\:& w\leq 0
\end{cases}
\end{equation}
(for any undirected pair of nodes) remains naturally induced. As shown in \cite{Parisi2020}, the problem \eqref{eq:nlp} has to be slightly generalized; still, its argument for the specific network $\mathbf{W}^*$ becomes

\begin{equation}
\mathcal{G}_{\text{CReM}}=-\sum_{i=1}^N\theta_is_i(\mathbf{W}^{*})+\sum_{i=1}^N\sum_{\substack{j=1\\(j<i)}}^Nf_{ij}\log\left[\theta_i+\theta_j\right]
\end{equation}
where the quantity $f_{ij}=\sum_\mathbf{A}P(\mathbf{A})a_{ij}$ represents the expected value of $a_{ij}$ over the ensemble of binary configurations defining the binary model taken as input (i.e. the marginal probability of an edge existing between nodes $i$ and $j$). It follows that the CReM first-order optimality conditions read

\begin{eqnarray}
\nabla_{\theta_i}\mathscr{L}_\mathrm{CReM}&=&-s_i(\mathbf{W}^*)+\sum_{j(\neq i)=1}^N\frac{f_{ij}}{\theta_i+\theta_j}\nonumber\\
&=&-s_i(\mathbf{W}^*)+\langle s_i\rangle=0,\quad i=1\dots N.\nonumber\\
\end{eqnarray}

\paragraph{Resolution of the CReM.} Newton's and the quasi-Newton method can still be implemented via the recipe defined in eq. \eqref{eq:gensol} (see Appendix A for the definition of the CReM Hessian).

As for the UECM and the DECM, the fixed-point recipe for solving the system of equations embodying the CReM transforms the set of consistency equations

\begin{equation}\label{eq:sys4}
\theta_i=\left[\frac{s_i(\mathbf{W}^*)}{\sum_{\substack{j=1\\(j\neq i)}}^N\left(\frac{f_{ij}}{1+\theta_j/\theta_i}\right)}\right]^{-1},\quad i=1\dots N
\end{equation}
into an iterative recipe of the usual form, i.e. by considering the parameters at the left hand side and at the right hand side, respectively at the $n$-th and at the $(n-1)$-th iteration. Although a reduced recipe can, in principle, be defined, an analogous observation to the one concerning the UECM and the DECM holds: the mathematical nature of the strengths (now, real numbers) increases their heterogeneity, in turn causing the CReM algorithm to be reducible even less than the `mixed' models defined by discrete weights.

The initialization of the iterative recipe for solving the CReM has been implemented in the usual threefold way. The first set of initial values reads $\theta_i^{(0)}=-\ln\left[\frac{s_i(\mathbf{W}^*)}{\sqrt{2W}}\right]$, $i=1\dots N$; the second one is a variant of the position above, reading $\theta_i^{(0)}=-\ln\left[\frac{s_i(\mathbf{W}^*)}{\sqrt{N}}\right]$; the third one, instead, prescribes to randomly draw the value of each parameter from the uniform distribution defined on the unit interval, i.e. $\theta_i^{(0)}\sim U(0,1)$, $\forall\:i$.\\

When considering directed networks, the conditional probability distribution defining the CReM reads

\begin{equation}
q_{ij}(w_{ij}|a_{ij}=1)=
\begin{cases}
(\alpha_i+\beta_j)e^{-(\alpha_i+\beta_j)w}\: & w>0 \\
\:0\:& w\leq 0
\end{cases}
\end{equation}
for any two nodes $i$ and $j$ such that $i\neq j$; the set of equations \eqref{eq:sys4} can be generalized as follows

\begin{eqnarray}
\alpha_i=\left[\frac{s_i^{out}(\mathbf{W}^*)}{\sum_{\substack{j=1\\(j\neq i)}}^N\left(\frac{f_{ij}}{1+\beta_j/\alpha_i}\right)}\right]^{-1},\quad i=1\dots N\nonumber\\
\beta_i=\left[\frac{s_i^{in}(\mathbf{W}^*)}{\sum_{\substack{j=1\\(j\neq i)}}^N\left(\frac{f_{ji}}{1+\alpha_j/\beta_i}\right)}\right]^{-1},\quad i=1\dots N\nonumber\\
\end{eqnarray}
and analogously for the sets of values initializing them.\\

\begin{table*}[t!]
\centering
\begin{tabular}{|l|c|c||c|c|c||c|c|c|
|c|c|c|}
\hline
\multicolumn{3}{c}{} & \multicolumn{3}{c}{Newton} & \multicolumn{3}{c}{Quasi-Newton} & \multicolumn{3}{c}{Fixed-point} \\
\hline
& \textit{N} & \textit{L} & MRDE & MRSE & Time (s) & MRDE & MRSE & Time (s) & MRDE & MRSE & Time (s)\\
\hline
e-MID 00 & 196 & 10618 & $\simeq 3\cdot 10^{-7}$ & $\simeq 2 \cdot 10^{-10} $& $\simeq 0.9 $ & $\simeq 5 \cdot 10^{-7}$ & $\simeq 2 \cdot 10^{-6}$ & $\simeq 0.9$ & $\simeq 4 \cdot 10^{-14}$ & $\simeq 8\cdot 10^{-5}$ & $\simeq 0.09$ \\
\hline
e-MID 01 & 185 & 8951 & $\simeq 3 \cdot 10^{-15}$ & $\simeq 1 \cdot 10^{-10}$ & $\simeq 0.9$ & $\simeq 7 \cdot 10^{-10}$ & $\simeq 5 \cdot 10^{-6}$ & $\simeq 1$ & $\simeq 7 \cdot 10^{-9}$ & $\simeq  1 \cdot 10^{-4}$ & $\simeq 0.12$ \\
\hline
e-MID 02 & 177 & 7252 & $\simeq 6 \cdot 10^{-9}$ & $\simeq 2 \cdot 10^{-13}$ & $\simeq 0.7$ & $\simeq 3 \cdot 10^{-8}$ & $\simeq 5 \cdot 10^{-6}$ & $\simeq 1$  & $\simeq 8 \cdot 10^{-6}$ & $\simeq 3 \cdot 10^{-1}$ & $\simeq 0.08$ \\
\hline
e-MID 03 & 179 & 6814 & $\simeq 1 \cdot 10^{-12}$ & $\simeq 2 \cdot 10^{-10}$ & $\simeq 0.8$ & $\simeq 3 \cdot 10^{-9} $ & $\simeq 3 \cdot 10^{-3}$ & $\simeq 0.7$ & $\simeq 4 \cdot 10^{-15} $ & $\simeq 8 \cdot 10^{-4}$ & $\simeq 0.1$ \\
\hline
e-MID 04 & 180 & 6136 & $\simeq 9 \cdot 10^{-10}$ & $\simeq 2 \cdot 10^{-7}$ & $\simeq 0.9 $ & $\simeq 5 \cdot 10^{-15} $ & $\simeq 8 \cdot 10^{-5}$ & $\simeq 0.8$ & $\simeq 5\cdot 10^{-9} $ & $\simeq 6 \cdot 10^{-4} $ & $\simeq 0.13$ \\
\hline
e-MID 05 & 176 & 6203 & $\simeq 3 \cdot 10^{-10}$ & $\simeq 5 \cdot 10^{-9}$ & $\simeq 0.7$ & $\simeq 2 \cdot 10^{-14}$ & $\simeq 2 \cdot 10^{-3}$ & $ \simeq 0.7$ & $\simeq 5 \cdot 10^{-9}$ & $\simeq 1\cdot 10^{-3}$ & $\simeq 0.2$ \\
\hline
e-MID 06 & 177 & 6132  & $\simeq 1 \cdot 10^{-10}$ & $\simeq 7 \cdot 10^{-12}$ & $\simeq 0.7$ & $\simeq 2 \cdot 10^{-13}$ & $\simeq 3 \cdot 10^{-3}$ & $\simeq 0.8$ & $\simeq 8 \cdot 10^{-11}$ & $\simeq 5 \cdot 10^{-1}$ & $\simeq 0.14$ \\
\hline
e-MID 07 & 178 & 6330 & $\simeq 3 \cdot 10^{-6}$ & $\simeq 3 \cdot 10^{-13}$ & $\simeq 1$ & $\simeq 1 \cdot 10^{-7}$ & $\simeq 8 \cdot 10^{-6}$ & $ \simeq 1.2 $ & $\simeq 7 \cdot 10^{-12}$ & $\simeq 7 \cdot 10^{-1}$ & $\simeq 0.14$ \\
\hline
e-MID 08 & 173 & 4767 & $\simeq 3 \cdot 10^{-10}$ & $\simeq 8 \cdot 10^{-13}$ & $\simeq 0.8$ & $\simeq 1 \cdot 10^{-9}$ & $\simeq 1 \cdot 10^{-3}$ & $\simeq 0.7$ & $\simeq 3 \cdot 10^{-9}$ & $\simeq 8 \cdot 10^{-1}$ & $ \simeq0.07$ \\
\hline
e-MID 09 & 156 & 2961 & $\simeq 4 \cdot 10^{-11}$ & $\simeq 3 \cdot 10^{-12}$ & $\simeq 0.6$ & $\simeq 1 \cdot 10^{-7}$ & $\simeq 9 \cdot 10^{-5}$ & $\simeq 0.7$ & $\simeq 8 \cdot 10^{-10}$ & $\simeq 2 \cdot 10^{-3}$ & $ \simeq 0.11$ \\
\hline
e-MID 10 & 135 & 2743 & $\simeq 2 \cdot 10^{-11}$ & $\simeq 2 \cdot 10^{-9}$ & $\simeq 0.7$ & $\simeq 7 \cdot 10^{-13}$ & $\simeq 5 \cdot 10^{-5}$ & $\simeq 0.5$ & $\simeq 5 \cdot 10^{-9}$ & $\simeq 2 \cdot 10^{-1}$ & $\simeq 0.05$ \\
\hline
\end{tabular}
\caption{Performance of Newton's, quasi-Newton and the fixed-point algorithm to solve the system of equations defining the (directed version of the) CReM, on a set of real-world WDNs. All algorithms stop because the condition $||\nabla\mathscr{L}(\vec{\theta})||_2\leq 10^{-8}$ is satisfied. For what concerns accuracy, the two most accurate methods are Newton's and the quasi-Newton one; for what concerns speed, the fastest method is the fixed-point one. Only the results corresponding to the best choice of initial conditions are reported.}
\label{tab:dircrem}
\end{table*}

\paragraph{Rescaling the CReM algorithm.} Although the equations defining the CReM algorithm cannot be effectively reduced, they can be opportunely \textit{rescaled}. To this aim, let us consider directed configurations and the system

\begin{equation}\label{sysres}
\left\{\begin{array}{ll}
\sum_{\substack{j=1\\j(\neq i)}}^N\frac{f_{ij}}{\alpha_i(\kappa)+\beta_j(\kappa)}&=\frac{s_i^{out}(\mathbf{W}^*)}{\kappa},\quad\forall\:i\\
\sum_{\substack{j=1\\j(\neq i)}}^N\frac{f_{ji}}{\alpha_j(\kappa)+\beta_i(\kappa)}&=\frac{s_i^{in}(\mathbf{W}^*)}{\kappa},\quad\forall\:i
\end{array}\right.
\end{equation}
where the sufficient statistics has been divided by an opportunely defined factor (in this case, $\kappa$) and the symbols $\alpha_i(\kappa)$, $\alpha_j(\kappa)$, $\beta_i(\kappa)$ and $\beta_j(\kappa)$ stress that the solution we are searching for is a function of the parameter $\kappa$ itself. In fact, a solution of the system above reads 

\begin{eqnarray}\label{sysres2}
\alpha^*_i&=&\frac{\alpha^*_i(\kappa)}{\kappa},\quad\forall\:i\\
\beta^*_i&=&\frac{\beta^*_i(\kappa)}{\kappa},\quad\forall\:i
\end{eqnarray}
as it can be proven upon substituting it back into eqs. (\ref{sysres}) and noticing that $\{\alpha^*_i\}_{i=1}^N$ and $\{\beta^*_i\}_{i=1}^N$ are solutions of the system of equations defined in (\ref{sysres}). As our likelihood maximization problem admits a unique, global maximum, the prescription above allows us to easily identify it. Rescaling will be tested in order to find out if our algorithms are enhanced by it under some respect (e.g. accuracy or speed).\\

\paragraph{Performance testing.} Before commenting on the performance of the three algorithms in solving the system of equations defining the CReM, let us stress once more that the formulas presented so far are perfectly general, working for any binary recipe one may want to employ. In what follows, we will test the CReM by posing $f_{ij}\equiv p_{ij}^\text{UBCM}$ and $f_{ij}\equiv p_{ij}^\text{DBCM}$.\\

To test the effectiveness of our algorithms in solving the CReM on undirected networks we have considered the synaptic network of the worm \emph{C. Elegans} \citep{Oshio2003} and the eight daily snapshots of the Bitcoin Lightning Network \citep{Lin2020}; the directed version of the CReM has, instead, been solved on the Electronic Italian Interbank Market (e-MID) during the decade 2000-2010 \cite{Iori2006}. Before commenting on the results of our numerical exercises, let us, first, describe how the latter ones have been carried out.\\

As for the discrete `mixed' models, the accuracy of each algorithm in reproducing the constraints defining the CReM has been quantified via the Maximum Relative Degree Error and the Maximum Relative Strength Error metrics, whose definition is provided by eqs. (\ref{eq:errsa}), (\ref{eq:errsb}) and (\ref{eq:errs2a}), (\ref{eq:errs2b}) for the undirected and the directed case, respectively. Analogously, the three `stop criteria' for each algorithm are the same ones that we have adopted for the other models (and consist in a condition on the Euclidean norm of the gradient of the likelihood function, i.e. $||\nabla\mathscr{L}(\vec{\theta})||_2\leq 10^{-8}$, a condition on the Euclidean norm of the vector of differences between the values of the parameters at subsequent iterations, i.e. $||\Delta\vec{\theta}||_2\leq 10^{-8}$, and a condition on the maximum number of iterations, i.e. 10000 steps).\\

The results about the performance of our three algorithms are reported in \textcolor{black}{Table \ref{tab:crem}} and in \textcolor{black}{Table \ref{tab:dircrem}}. Let us start by commenting the results reported in \textcolor{black}{Table \ref{tab:crem}} and concerning undirected networks. Generally speaking, Newton's method is the most accurate one (its largest maximum errors span intervals, across all configurations, that amount at $10^{-11}\lesssim\text{MRDE}_\text{Newton}\lesssim 10^{-8}$ and $10^{-5}\lesssim\text{MRSE}_\text{Newton}\lesssim 10^{-4}$) although it scales very badly with the size of the network on which it is tested (the amount of time, measured in seconds, required by it to achieve convergence spans an interval, across all configurations, that amounts at $0.08\leq T_\text{Newton}^\text{reduced}\leq 1188)$.

The quasi-Newton method, on the other hand, is very accurate on the degrees (as already observed in the UBCM case) but not so accurate in reproducing the weighted constraints (its largest maximum errors span intervals, across all configurations, that amount at $\text{MRDE}_\text{Quasi-Newton}\simeq 10^{-7}$ and $10^{-6}\lesssim\text{MRSE}_\text{Quasi-Newton}\lesssim 6$). Moreover, it scales even worse than Newton's method with the size of the network on which it is tested (the amount of time, measured in seconds, required by it to achieve convergence spans an interval, across all configurations, that amounts at $5\leq T_\text{Quasi-Newton}\leq 15888$).

The performance of the fixed-point recipe is, somehow, intermediate between that of Newton's and the quasi-Newton method. For what concerns accuracy, it is more accurate in reproducing the binary constraints than in reproducing the weighted ones (its largest maximum errors span intervals, across all configurations, that amount at $\text{MRDE}_\text{fixed-point}\simeq 10^{-9}$ and $10^{-8}\lesssim\text{MRSE}_\text{fixed-point}\lesssim 10^{-1}$) although it outperforms Newton's method, sometimes. For what concerns scalability, the fixed-point method is the less sensitive one to the growing size of the considered configurations: hence, it is also the fastest one (the amount of time, measured in seconds, required by it to achieve convergence spans an interval, across all configurations, that amounts at $0.01\leq T_\text{fixed-point}\leq 550$).

Moreover, while Newton's and the fixed-point method stop because the condition on the norm of the likelihood is satisfied, the quasi-Newton method is often found to satisfy the limit condition on the number of steps (i.e. it runs for 10000 steps and, then, stops).

Interestingly, the fact that the CReM cannot be reduced (at least not to a comparable extent with the one characterizing purely binary models) reveals a dependence on the network size of Newton's and the quasi-Newton algorithms. The reason may lie in the evidence that both Newton's and the quasi-Newton method require (some proxy of) the Hessian matrix of the system of equations defining the CReM to update the value of the parameters: as already observed, the order of the latter - which is $O(N^2)$ for Newton's method and $O(N)$ for the quasi-Newton one - can make its calculation (very) time demanding.\\

Let us now move to comment on the performance of our algorithms when applied to solve the directed version of the CReM (see \textcolor{black}{Table \ref{tab:dircrem}}). Overall, all methods perform much better than in the undirected case, stopping because the condition on the norm of the likelihood is satisfied.

In fact, all of them are very accurate in reproducing the purely binary constraints, their largest maximum errors spanning intervals, across all configurations, that amount at $10^{-12}\lesssim\text{MRDE}_\text{Newton}\lesssim 10^{-6}$, $10^{-14}\lesssim\text{MRDE}_\text{Quasi-Newton}\lesssim 10^{-6}$ and $10^{-15}\lesssim\text{MRDE}_\text{fixed-point}\lesssim 10^{-8}$; for what concerns the weighted constraints, instead, the two most accurate methods are Newton's and the quasi-Newton one, their largest maximum errors spanning intervals, across all configurations, that amount at $10^{-13}\lesssim\text{MRSE}_\text{Newton}\lesssim 10^{-7}$ and $10^{-6}\lesssim\text{MRSE}_\text{Quasi-Newton}\lesssim 10^{-3}$ (the fixed-point method performs worse than them, since $10^{-3}\lesssim\text{MRSE}_\text{fixed-point}\lesssim 10^{-1}$).

For what concerns speed, the amount of time, measured in seconds, required by Newton's, the quasi-Newton and the fixed-point algorithms to achieve convergence spans an interval, across all configurations, that amounts at $0.6\leq T_\text{Newton}^\text{reduced}\leq 1$, $0.5\leq T_\text{Quasi-Newton}^\text{reduced}\leq 1.2$ and $0.05\leq T_\text{fixed-point}^\text{reduced}\leq 0.2$, respectively. Hence, all methods are also very fast - the fixed-point being systematically the fastest one.

As already stressed above, the fact that the e-MID number of nodes remains approximately constant throughout the considered time interval masks the strong dependence of the performance of Newton's and the quasi-Newton method on the network size.

Lastly, while rescaling the system of equations defining the CReM improves neither the accuracy nor the speed of any of the three algorithms considered here, differences in their speed of convergence, caused by the choice of a particular set of initial conditions, are observable: the `uniform' prescription outperforms the other ones (for both the undirected and the directed version of the CReM).\\

As usual, let us comment on the algorithm to sample the CReM ensemble - for the sake of simplicity, on the undirected cae. As for the UECM it can be compactly achieved by implementing a two-step procedure, the only difference lying in the functional form of the distribution from which weights are sampled. Given a specific pair of vertices $i,j$ (with $i<j$), the first link can be drawn from a Bernoulli distribution with probability $p^\text{UBCM}_{ij}$ while the remaining $(w-1)$ ones can be drawn from an exponential distribution whose parameter reads $\theta_i+\theta_j$. The computational complexity of the sampling process is, again, $O(N^2)$ and the pseudo-code for explicitly sampling the CReM ensemble is summed up by Algorithm 6.

\begin{algorithm}[H]
\begin{algorithmic}[1]
\For{$m=1\dots |E|$}
\State $\mathbf{W}=\mathbf{0};$
\For{$i=1\dots N$}
\For{$j=1\dots N$ \textrm{and} $j<i$}
\If{$\text{RandomUniform}[0,1]\leq p_{i\alpha}^\text{UBCM}$}
\State $w_{ij}=w_{ji}=\text{RandomExponential}[\theta_i+\theta_j];$
\Else
\State $w_{ij}=w_{ji}=0;$
\EndIf
\EndFor
\EndFor
\State $\text{Ensemble}[m]=\mathbf{W};$
\EndFor
\caption{Sampling the CReM ensemble}
\label{code:crem}
\end{algorithmic}
\end{algorithm}

\section{Discussion}\label{results}

The exercises carried out so far have highlighted a number of (stylized) facts concerning the performance of the three algorithms tested: in what follows, we will briefly sum them up.\\

\paragraph{Newton's method.} Overall, Newton's method is very accurate - often, the most accurate one - in reproducing both the binary and the weighted constraints; moreover, it represent the only viable alternative when the most complicated models are considered (i.e. the UECM and the DECM, respectively defined by a system of $2N$ and $4N$ coupled, non-linear equations). However, the time required to run Newton's method on a given model seems to be quite dependent on the network size, especially whenever the corresponding system of equations cannot be reduced - see the case of the undirected CReM, run on the Bitcoin Lightning Network. Since one of the reasons affecting the bad scaling of Newton's method with the network size is the evaluation of the Hessian matrix defining a given model, this algorithm has to be preferred for largely reducible networks.\\

\paragraph{Quasi-Newton method.} For all the networks considered here, the quasi-Newton method we have implemented is nothing else than the diagonal version of the traditional Newton's method. Even if this choice greatly reduces the number of entries of the Hessian matrix which are needed (i.e. just $N$ elements for the undirected version of the CReM, $2N$ elements for the UECM and the directed version of the CReM and $4N$ elements for the DECM) dimensionality may still represent an issue to achieve fast convergence. Moreover, since the diagonal approximation of the Hessian matrix is not necessarily always a good one, the quasi-Newton method may require more time than Newton's one to achieve the same level of accuracy in reproducing the constraints. However, when such an approximation is a good one, the `regime' in which the quasi-Newton method outperforms the competitors seems to be the one of small, non-reducible networks (e.g. see the results concerning the DBCM run on the WTW) - althogh, in cases like these, Newton's method may still be a strong competitor.\\

\paragraph{Fixed-point method.} From a purely theoretical point of view, the fixed-point recipe is the fastest one, since the time required to evaluate the generic $n$-th step is (only) due to the evaluation of the model-specific map at the $(n-1)$-th iteration. Strictly speaking, however, this holds true for a single step: if the number of steps required for convergence is large, in fact, the total amount of time required by the fixed-point method can be large as well. Overall, however, this algorithm has to be preferred for large, non-reducible networks: this is the case of the (undirected version of the) CReM, run on the $8$-th snapshot of the Bitcoin Lightning Network (i.e. day 17-07-19) and requiring a little more than one minute to achieve an accuracy of $\text{MRDE}_\text{fixed-point}\gtrsim 10^{-10}$ and $\text{MRSE}_\text{fixed-point}\simeq 10^{-1}$; naturally, the method is not as accurate as Newton's one, for which $\text{MRDE}_\text{Newton}\gtrsim 10^{-12}$ and $\text{MRSE}_\text{Newton}\simeq 10^{-6}$ but is much faster as Newton's algorithm requires $\simeq 1188$ seconds to converge.\\

\paragraph{The `NEMTROPY' Python package.} As an additional result, we release a comprehensive package, coded in Python, that implements the three aforementioned algorithms on all the ERGMs considered in the present work. Its name is `NEMTROPY', an acronym standing for `Network Entropy Maximization: a Toolbox Running On Python', and is freely downloadable at the following URL: \texttt{https://pypi.org/project/NEMtropy/}.\\

Alternative techniques to improve accuracy and speed have been tested as well, as the one of coupling two of the algorithms considered above. In particular, we have tried to solve the (undirected version of the) CReM by running the fixed-point algorithm and using the solution of the latter as input for the quasi-Newton method. The results are reported in \textcolor{black}{Table \ref{tab:crem2}}: as they clearly show, the coupled algorithm is indeed more accurate that the single methods composing it and much faster than the quasi-Newton one (for some snapshots, more accurate and even faster than Newton's method). Techniques like these are, in general, useful to individuate better initial conditions than the completely random ones: a first run of the fastest method may be, in fact, useful to direct the most accurate algorithm towards the (best) solution. This is indeed the case, upon considering that the quasi-Newton method, now, stops because the condition $||\nabla\mathscr{L}(\vec{\theta})||_2\leq 10^{-8}$ is satisfied - and not for having reached the limit of 10000 steps.\\

\begin{table*}[t!]
\centering
\begin{tabular}{|l|c|c|c|c|}
\hline
\multicolumn{1}{c}{} & \multicolumn{4}{c}{Fixed-point + Quasi-Newton}\\
\hline
& MRDE & MADE & MRSE & Time (s) \\
\hline
BLN 24-01-18 &
$\simeq 2 \cdot 10^{-9}$ & $\simeq 1.1 \cdot 10^{-7}$ & $\simeq 1 \cdot 10^{-5}$ & $\simeq 0.1 $\\
\hline
BLN 25-02-18 &
$\simeq 1.3 \cdot 10^{-9}$ & $\simeq 1.5 \cdot 10^{-6}$ & $\simeq 1 \cdot 10^{-5}$ & $\simeq 1.6 $\\
\hline
BLN 30-03-18 &
$\simeq 1 \cdot 10^{-9}$ & $\simeq 1.3 \cdot 10^{-7}$ & $\simeq 7.8 \cdot 10^{-7}$ & $\simeq 2.2 $\\
\hline
BLN 13-07-18 &
$\simeq 7.5 \cdot 10^{-10}$ & $\simeq 4.2 \cdot 10^{-4}$ & $\simeq 5.3 \cdot 10^{-5}$ & $\simeq 200 $\\
\hline
BLN 19-12-18 &
$\simeq 7.5 \cdot 10^{-10}$ & $ \simeq 1.7 \cdot 10^{-8}$ & $ \simeq 1.1 \cdot 10^{-9}$ & $\simeq 7 $\\
\hline
BLN 30-01-19 &
$\simeq 6.2 \cdot 10^{-10}$ & $\simeq 1.8 \cdot 10^{-5}$ & $ \simeq 4.1 \cdot 10^{-6}$ & $\simeq 614 $\\
\hline
BLN 01-03-19 &
$\simeq 5.7 \cdot 10^{-10}$ & \textbf{$\simeq 5.4 \cdot 10^{-6}$} & $ \simeq 9.1 \cdot 10^{-6}$ & $\simeq 961 $\\
\hline
BLN 17-07-19 &
$\simeq 4.9 \cdot 10^{-10}$ & $ \simeq 1.3 \cdot 10^{-3}$ & $ \simeq 3.5 \cdot 10^{-3}$ & $\simeq 3350 $\\
\hline
\end{tabular}
\caption{Performance of the algorithm coupling fixed-point and quasi-Newton to solve the system of equations defining the (undirected version of the) CReM, on a set of real-world WUNs. The algorithm stops because the condition $||\nabla\mathscr{L}(\vec{\theta})||_2\leq 10^{-8}$ is satisfied. As the results reveal, it is more accurate that the single methods composing it and much faster than the quasi-Newton one - for some snapshots, more accurate and even faster than Newton's method. Only the results corresponding to the best choice of initial conditions are reported.}
\label{tab:crem2}
\end{table*}

We would like to end the discussion about the results presented in this contribution by explicitly mentioning a circumstance that is frequently met when studying economic and financial networks. When considering systems like these, the information about the number of neighbours of each node is typically not accessible: as a consequence, the models constraining both binary and weighted information cannot be employed as they have presented in this contribution.

Alternatives exist and rest upon the existence of some kind of relationship between binary and weighted constraints. In the case of undirected networks, such a relationship is usually written as

\begin{equation}
e^{-\theta_i}=\sqrt{z}s_i
\end{equation}
and establishes that the Lagrange multipliers controlling for the degrees are linearly proportional to the strengths. If this is the case (or a valid reason exists for this to be the case), the expression for the probability that any two nodes are connected becomes 

\begin{equation}
p_{ij}^\text{dcGM}=\frac{zs_is_j}{1+zs_is_j}\quad\forall\:i<j
\end{equation}
the acronym standing for \textit{degree-corrected Gravity Model} \cite{Cimini2015a}. The (only) unknown parameter $z$ must be numerically estimated by employing some kind of topological information; this is usually represented by (a proxy of) the network link density, used to instantiate the (only) likelihood condition

\begin{equation}
L(\mathbf{A}^*)=\langle L\rangle=\sum_{i=1}^N\sum_{\substack{j=1\\(j<i)}}^N\frac{zs_is_j}{1+zs_is_j};
\end{equation}
once the equation above has been solved, the set of coefficients $\{p_{ij}^\text{dcGM}\}_{i,j=1}^N$ can be either employed 1) to, first, estimate the degrees and, then, solve the UECM \cite{Cimini2015b} or 2) within the CReM framework, via the identification $f_{ij}\equiv p_{ij}^\text{dcGM}$, to estimate the parameters controlling for the weighted constraints.

\section{Conclusions}\label{conclusions}

The definition and correct implementation of null models is a crucial issue in network analysis: the present contribution focuses on (a subset of) the ones constituting the so-called ERG framework - a choice driven by the evidence that they are the most commonly employed ones for tasks as different as network reconstruction, pattern detection, graph enumeration. The optimization of the likelihood function associated to them is, however, still problematic since it involves the resolution of large systems of coupled, non-linear equations. 

Here, we have implemented and compared three algorithms for numerical optimization, with the aim of finding the one performing best (i.e. being both accurate and fast) for each model. What emerges from our results is that there is no a unique method which is both accurate and fast for all models on all configurations: under this respect, performance is just a trade-off between accuracy and speed. However, some general conclusions can still be drawn.

Newton's method is the one requiring the largest amount of information per step (in fact, the entire Hessian matrix has to be evaluated at each iteration): hence, it is the most accurate one but, for the same reason, often the one characterized by the slowest performance. A major drawback of Newton's method is that of scaling very badly with the network size.

At the opposite extreme lies the fixed-point algorithm, theoretically the fastest one but, often, among the least accurate ones (at least, for what concerns the weighted constraints); the performance if the quasi-Newton method often lies in-between the performances of the two methods above, by achieving an accuracy that is larger than the one achieved by the fixed-point algorithm, requiring less time that the one needed by Newton's method.

Overall, while Newton's method seems to perform best on either relatively small or greatly reducible networks, the fixed-point method must be preferred for large, non-reducible configurations. Deviations from this (over simplified) picture are, however, clearly visible.\\

Future work concerns the application of the aforementioned three numerical recipes to the models that have not found place here. For what concerns the set of purely binary constraints, the ones defining the \textit{Reciprocal Binary Configuration Model} (RBCM) \cite{Squartini2011} deserve to be mentioned.

For what concerns the `mixed' constraints, instead, the CReM framework is versatile enough \textcolor{black}{to accommodate a quite large number of variants. In the present work, we have `limited' ourselves to combine the UBCM and the DBCM with (conditional) distributions of \emph{continuous} weights: a first, obvious, generalization is that of considering the \emph{discrete} versions of such models, defined by the positions}

\begin{equation}
\langle w_{ij}\rangle_\text{d-CReM}^\text{und}=\frac{f_{ij}}{1-e^{-\beta_j-\beta_i}}
\end{equation}
\textcolor{black}{with $f_{ij}\equiv p_{ij}^\text{UBCM}$ and}

\begin{equation}
\langle w_{ij}\rangle_\text{d-CReM}^\text{dir}=\frac{f_{ij}}{1-e^{-\gamma_j-\delta_i}}
\end{equation}
\textcolor{black}{with $f_{ij}\equiv p_{ij}^\text{DBCM}$; a second one concerns the continuous versions of the UECM and of the DECM, respectively defined by the positions \cite{Gabrielli2019}}

\begin{equation}
p_{ij}^{\text{UECM}}=\frac{e^{-\alpha_i-\alpha_j}}{\beta_i+\beta_j+e^{-\alpha_i-\alpha_j}}
\end{equation}
\textcolor{black}{and}

\begin{equation}
p_{ij}^{\text{DECM}}=\frac{e^{-\alpha_i-\beta_j}}{\gamma_i+\delta_j+e^{-\alpha_i-\beta_j}}.
\end{equation}

\section{Authors contributions}

FS, GC, MZ, TS developed the methods and designed the research. NV, MB, EM, GT performed the analysis (NV: DBCM, DECM, RBCM; MB: BiCM; EM: UBCM, UECM, CReM; GT: preliminary version of the BiCM). FS, GC, MZ, TS wrote the manuscript. All authors reviewed and approved the manuscript.

\section{Acknowledgements}

FS and TS acknowledge support from the European project SoBigData++ (GA. 871042). GC and MZ acknowledge support from the PAI project ORCS (`Optimized Reconstruction of Complex networkS'), funded by the IMT School for Advanced Studies Lucca. FS also acknowledges support from the Italian `Programma di Attivit\`a Integrata' (PAI) project `TOol for Fighting FakEs' (TOFFE), funded by IMT School for Advanced Studies Lucca.

\section*{Appendix A: computing\\the Hessian matrix}\label{appa}

As we showed in the main text, the Hessian matrix of our likelihood function is `minus' the covariance matrix of the constraints, i.e.

\begin{equation}
H_{ij}=\frac{\partial^2\mathscr{L}(\vec{\theta})}{\partial\theta_i\partial\theta_j}=-\text{Cov}[C_i,C_j],\quad i,j=1\dots M;
\end{equation}
interestingly, a variety of alternative methods exists to explicitly calculate the generic entry $H_{ij}$, i.e. 1) taking the second derivatives of the likelihood function characterizing the method under analysis, 2) taking the first derivatives of the expectation values of the constraints characterizing the method under analysis, 3) calculating the moments of the pair-specific probability distributions characterizing each method.

\subsection{UBCM: binary undirected graphs\\with given degree sequence}

The Hessian matrix for the UBCM is an $N\times N$ symmetric table with entries reading

\begin{equation}
H_\text{UBCM}=
\begin{cases}
\text{Var}[k_i]=\sum_{\substack{j=1\\(j\neq i)}}^Np_{ij}(1-p_{ij}),\quad \forall\:i\\
\text{Cov}[k_i,k_j]=p_{ij}(1-p_{ij}),\quad \forall\:i\neq j\\
\end{cases}
\end{equation}
where $p_{ij}\equiv p_{ij}^\text{UBCM}$. Notice that $\text{Var}[k_i]=\sum_{\substack{j=1\\(j\neq i)}}^N\text{Cov}[k_i,k_j],\:\forall\:i$.

\subsection{DBCM: binary directed graphs\\with given in-degree and out-degree sequences}

The Hessian matrix for the DBCM is a $2N\times 2N$ symmetric table that can be further subdivided into four $N\times N$ blocks whose entries read

\begin{equation}
H_\text{DBCM}=
\begin{cases}
\text{Var}[k_i^{out}]=\sum_{\substack{j=1\\(j\neq i)}}^Np_{ij}(1-p_{ij}),\quad \forall\:i\\
\text{Var}[k_i^{in}]=\sum_{\substack{j=1\\(j\neq i)}}^Np_{ji}(1-p_{ji}),\quad \forall\:i\\
\text{Cov}[k_i^{out},k_j^{in}]=p_{ij}(1-p_{ij}),\quad \forall\:i\neq j\\
\text{Cov}[k_j^{out},k_i^{in}]=p_{ji}(1-p_{ji}),\quad \forall\:i\neq j\\
\end{cases}
\end{equation}
while $\text{Cov}[k_i^{out},k_i^{in}]=\text{Cov}[k_i^{out},k_j^{out}]=\text{Cov}[k_i^{in},k_j^{in}]=0$ and $p_{ij}\equiv p_{ij}^\text{DBCM}$.\\

Notice that the Hessian matrix of the BiCM mimicks the DBCM one, the only difference being that the probability coefficients are now indexed by $i$ and $\alpha$: for example, in the BiCM case, one has that $\text{Cov}[k_i,d_\alpha]=p_{i\alpha}(1-p_{i\alpha})$, $\forall\:i,\alpha$.

\subsection{UECM: weighted undirected graphs with given strengths and degrees}

The Hessian matrix for the UECM is a $2N\times 2N$ symmetric table that can be further subdivided into four blocks (each of which with dimensions $N\times N$). In order to save space, the expressions indexed by the single subscript $i$ will be assumed as being valid $\forall\:i$, while the ones indexed by a double subscript $i,j$ will be assumed as being valid $\forall\:i\neq j$. The entries of the diagonal blocks read

\begin{equation}
H_\text{UECM}=
\begin{cases}
\frac{\partial^2\mathscr{L}_\text{UECM}}{\partial\alpha_i^2}=\text{Var}[k_i]=\sum_{\substack{j=1\\(j\neq i)}}^Np_{ij}(1-p_{ij})\\
\frac{\partial^2\mathscr{L}_\text{UECM}}{\partial\alpha_i\alpha_j}=\text{Cov}[k_i,k_j]=p_{ij}(1-p_{ij})
\end{cases}
\end{equation}
and

\begin{equation}
H_\text{UECM}=
\begin{cases}
\frac{\partial^2\mathscr{L}_\text{UECM}}{\partial\beta_i^2}=\text{Var}[s_i]=\sum_{\substack{j=1\\(j\neq i)}}^N\frac{p_{ij}(1-p_{ij}+e^{-\beta_i-\beta_j})}{(1-e^{-\beta_i-\beta_j})^2}\\
\frac{\partial^2\mathscr{L}_\text{UECM}}{\partial\beta_i\beta_j}=\text{Cov}[s_i,s_j]=\frac{p_{ij}(1-p_{ij}+e^{-\beta_i-\beta_j})}{(1-e^{-\beta_i-\beta_j})^2}
\end{cases}
\end{equation}
where $p_{ij}\equiv p_{ij}^\text{UECM}$. On the other hand, the entries of the off-diagonal blocks read

\begin{equation}
H_\text{UECM}=
\begin{cases}
\frac{\partial^2\mathscr{L}_\text{UECM}}{\partial\alpha_i\partial\beta_i}=\text{Cov}[k_i,s_i]=\sum_{\substack{j=1\\(j\neq i)}}^N\frac{p_{ij}(1-p_{ij})}{1-e^{-\beta_i-\beta_j}}\\
\frac{\partial^2\mathscr{L}_\text{UECM}}{\partial\alpha_i\partial\beta_j}=\text{Cov}[k_i,s_j]=\frac{p_{ij}(1-p_{ij})}{1-e^{-\beta_i-\beta_j}}\\
\end{cases}
\end{equation}
with $p_{ij}\equiv p_{ij}^\text{UECM}$.

\subsection{DECM: weighted directed graphs\\with given strengths and degrees}

The Hessian matrix for the DECM is a $4N\times 4N$ symmetric table that can be further subdivided into four blocks (each of which with dimensions $N\times N$). As for the UECM, in order to save space, the expressions indexed by the single subscript $i$ will be assumed as being valid $\forall\:i$, while the ones indexed by a double subscript $i,j$ will be assumed as being valid $\forall\:i\neq j$. The entries of the diagonal blocks read

\begin{equation}
H_\text{DECM}=
\begin{cases}
\frac{\partial^2\mathscr{L}_\text{DECM}}{\partial\alpha_i^2}=\text{Var}[k_i^{out}]=\sum_{\substack{j=1\\(j\neq i)}}^Np_{ij}(1-p_{ij})\\
\frac{\partial^2\mathscr{L}_\text{DECM}}{\partial\alpha_i\alpha_j}=\text{Cov}[k_i^{out},k_j^{out}]=0
\end{cases}
\end{equation}
and

\begin{equation}
H_\text{DECM}=
\begin{cases}
\frac{\partial^2\mathscr{L}_\text{DECM}}{\partial\beta_i^2}=\text{Var}[k_i^{in}]=\sum_{\substack{j=1\\(j\neq i)}}^Np_{ji}(1-p_{ji})\\
\frac{\partial^2\mathscr{L}_\text{DECM}}{\partial\beta_i\beta_j}=\text{Cov}[k_i^{in},k_j^{in}]=0
\end{cases}
\end{equation}
and

\begin{equation}
H_\text{DECM}=
\begin{cases}
\frac{\partial^2\mathscr{L}_\text{DECM}}{\partial\gamma_i^2}=\text{Var}[s_i^{out}]=\sum_{\substack{j=1\\(j\neq i)}}^N\frac{p_{ij}(1-p_{ij}+e^{-\gamma_i-\delta_j})}{(1-e^{-\gamma_i-\delta_j})^2}\\
\frac{\partial^2\mathscr{L}_\text{DECM}}{\partial\gamma_i\gamma_j}=\text{Cov}[s_i^{out},s_j^{out}]=0
\end{cases}
\end{equation}
and

\begin{equation}
H_\text{DECM}=
\begin{cases}
\frac{\partial^2\mathscr{L}_\text{DECM}}{\partial\delta_i^2}=\text{Var}[s_i^{in}]=\sum_{\substack{j=1\\(j\neq i)}}^N\frac{p_{ji}(1-p_{ji}+e^{-\gamma_j-\delta_i})}{(1-e^{-\gamma_j-\delta_i})^2}\\
\frac{\partial^2\mathscr{L}_\text{DECM}}{\partial\delta_i\delta_j}=\text{Cov}[s_i^{in},s_j^{in}]=0
\end{cases}
\end{equation}
where $p_{ij}\equiv p_{ij}^\text{DECM}$. On the other hand, the entries of the off-diagonal blocks read

\begin{equation}
H_\text{DECM}=
\begin{cases}
\frac{\partial^2\mathscr{L}_\text{DECM}}{\partial\alpha_i\partial\beta_i}=\text{Cov}[k_i^{out},k_i^{in}]=0\\
\frac{\partial^2\mathscr{L}_\text{DECM}}{\partial\alpha_i\partial\beta_j}=\text{Cov}[k_i^{out},k_j^{in}]=p_{ij}(1-p_{ij})\\
\end{cases}
\end{equation}
and

\begin{equation}
H_\text{DECM}=
\begin{cases}
\frac{\partial^2\mathscr{L}_\text{DECM}}{\partial\alpha_i\partial\gamma_i}=\text{Cov}[k_i^{out},s_i^{out}]=\sum_{\substack{j=1\\(j\neq i)}}^N\frac{p_{ij}(1-p_{ij})}{1-e^{-\gamma_i-\delta_j}}\\
\frac{\partial^2\mathscr{L}_\text{DECM}}{\partial\alpha_i\partial\gamma_j}=\text{Cov}[k_i^{out},s_j^{out}]=0\\
\end{cases}
\end{equation}
and

\begin{equation}
H_\text{DECM}=
\begin{cases}
\frac{\partial^2\mathscr{L}_\text{DECM}}{\partial\alpha_i\partial\delta_i}=\text{Cov}[k_i^{out},s_i^{in}]=0\\
\frac{\partial^2\mathscr{L}_\text{DECM}}{\partial\alpha_i\partial\delta_j}=\text{Cov}[k_i^{out},s_j^{in}]=\frac{p_{ij}(1-p_{ij})}{1-e^{-\gamma_i-\delta_j}}\\
\end{cases}
\end{equation}
and

\begin{equation}
H_\text{DECM}=
\begin{cases}
\frac{\partial^2\mathscr{L}_\text{DECM}}{\partial\beta_i\partial\gamma_i}=\text{Cov}[k_i^{in},s_i^{out}]=0\\
\frac{\partial^2\mathscr{L}_\text{DECM}}{\partial\beta_i\partial\gamma_j}=\text{Cov}[k_i^{in},s_j^{out}]=\frac{p_{ji}(1-p_{ji})}{1-e^{-\gamma_j-\delta_i}}\\
\end{cases}
\end{equation}
and

\begin{equation}
H_\text{DECM}=
\begin{cases}
\frac{\partial^2\mathscr{L}_\text{DECM}}{\partial\beta_i\partial\delta_i}=\text{Cov}[k_i^{in},s_i^{in}]=\sum_{\substack{j=1\\(j\neq i)}}^N\frac{p_{ji}(1-p_{ji})}{1-e^{-\gamma_j-\delta_i}}\\
\frac{\partial^2\mathscr{L}_\text{DECM}}{\partial\beta_i\partial\delta_j}=\text{Cov}[k_i^{in},s_j^{in}]=0\\
\end{cases}
\end{equation}
and

\begin{equation}
H_\text{DECM}=
\begin{cases}
\frac{\partial^2\mathscr{L}_\text{DECM}}{\partial\gamma_i\partial\delta_i}=\text{Cov}[s_i^{out},s_i^{in}]=0\\
\frac{\partial^2\mathscr{L}_\text{DECM}}{\partial\gamma_i\partial\delta_j}=\text{Cov}[s_i^{out},s_j^{in}]=\frac{p_{ij}(1-p_{ij}+e^{-\gamma_i-\delta_j})}{(1-e^{-\gamma_i-\delta_j})^2}
\end{cases}
\end{equation}
with $p_{ij}\equiv p_{ij}^\text{DECM}$.

\subsection{Two-step models for\\undirected and directed networks}

The Hessian matrix for the undirected two-step model considered here is an $N\times N$ symmetric table reading

\begin{equation}
H_\text{CReM}^\text{und}=
\begin{cases}
\text{Var}[s_i]=\sum_{\substack{j=1\\(j\neq i)}}^N\frac{f_{ij}}{(\theta_i+\theta_j)^2},\quad \forall\:i\\
\text{Cov}[s_i,s_j]=\frac{f_{ij}}{(\theta_i+\theta_j)^2},\quad \forall\:i\neq j\\
\end{cases}
\end{equation}
where $f_{ij}$ is given. In the directed case, instead, the Hessian matrix for the two-step model considered here is a $2N\times 2N$ symmetric table that can be further subdivided into four $N\times N$ blocks whose entries read

\begin{equation}
H_\text{CReM}^\text{dir}
\begin{cases}
\text{Var}[s_i^{out}]=\sum_{\substack{j=1\\(j\neq i)}}^N\frac{f_{ij}}{(\alpha_i+\beta_j)^2},\quad \forall\:i\\
\text{Var}[s_i^{in}]=\sum_{\substack{j=1\\(j\neq i)}}^N\frac{f_{ji}}{(\alpha_j+\beta_i)^2},\quad \forall\:i\\
\text{Cov}[s_i^{out},s_j^{in}]=\frac{f_{ij}}{(\alpha_i+\beta_j)^2},\quad \forall\:i\neq j\\
\end{cases}
\end{equation}
while $\text{Cov}[s_i^{out},s_i^{in}]=\text{Cov}[s_i^{out},s_j^{out}]=\text{Cov}[s_i^{in},s_j^{in}]=0$ and $f_{ij}$ is given.

\section*{Appendix B: a note on\\the change of variables}

In all methods we will considered in the present work, the variable $\theta_i$ appears in the optimality conditions only through negative exponential functions: it is therefore tempting to perform the change of variable $x_i\equiv e^{-\theta_i}$. Although this is often performed in the literature, one cannot guarantee that the new optimization problem remains convex: in fact, simple examples can be provided for which convexity is lost. This has several consequences, e.g. 1) convergence to the global maximum is no longer guaranteed (since the existence of a global maximum is no longer guaranteed as well), 2) extra-care is needed to guarantee that the Hessian matrix $\mathbf{H}$ employed in our algorithms is negative definite. While problem 2) introduces additional complexity only for Newton's method, problem 1) is more serious from a theoretical point of view.

Let us now address problem 1) in more detail. First, it is possible to prove that any stationary point for $\mathscr{L}(\vec{x})$ satisfies the optimality conditions for $\mathscr{L}(\vec{\theta})$ as well. In fact, the application of the `chain rule' leads to recover the set of relationships

\begin{equation}
\frac{\partial\mathscr{L}(\vec{\theta})}{\partial\theta_i}=\frac{\partial x_i}{\partial\theta_i}\frac{\partial\mathscr{L}(\vec{x})}{\partial x_i}=-x_i\frac{\partial\mathscr{L}(\vec{x})}{\partial x_i},\quad i=1\dots M;
\end{equation}
notice that requiring $\nabla_{\theta_i}\mathscr{L}(\vec{\theta})=0$ leads to require that either $\nabla_{x_i}\mathscr{L}(\vec{x})=0$ or $x_i=0$. As the second eventuality precisely identifies \textit{isolated} nodes (i.e. the nodes for which the constraint $C_i(\mathbf{G}^*)$, controlled by the multiplier $\theta_i$, is 0), one can get rid of it by explicitly removing the corresponding addenda from the likelihood function.

For what concerns convexity, let us explicitly calculate the Hessian matrix for the set of variables $\{x_i\}_{i=1}^M$. In formulas,

\begin{eqnarray}
\frac{\partial^2\mathscr{L}(\vec{x})}{\partial x_i^2}&=&e^{2\theta_i}\left(\frac{\partial^2\mathscr{L}(\vec{\theta})}{\partial\theta_i^2}+\frac{\partial \mathscr{L}(\vec{\theta})}{\partial\theta_i}\right),\quad i=1\dots M,\nonumber\\
\frac{\partial^2\mathscr{L}(\vec{x})}{\partial x_i\partial x_j}&=&e^{\theta_i+\theta_j}\left(\frac{\partial^2 \mathscr{L}(\vec{\theta})}{\partial\theta_i\partial\theta_j}\right),\quad \forall\:i\neq j\nonumber\\
\end{eqnarray}
according to the `chain rule' for second-order derivatives. More compactly,

\begin{equation}
\mathbf{H}_{\mathscr{L}(\vec{x})}=e^\Theta\circ\left(-\text{Cov}[C_i,C_j]+\mathbf{I}\cdot\nabla_{\vec{\theta}}\mathscr{L}(\vec{\theta})\right)
\end{equation}
where $\mathbf{I}$ is the identity matrix, the generic entry of the matrix $e^\Theta$ reads $\left[e^\Theta\right]_{ij}\equiv e^{\theta_i+\theta_j}$, $\forall\:i,j$ and the symbol `$\circ$' indicates the Hadamard (i.e. element-wise) product of matrices. In general, the expression above defines an \textit{indefinite} matrix, i.e. a neither positive nor negative (semi)definite one.

\section*{Appendix C: fixed point method\\in the multivariate case}

Equation \eqref{eq:fp} can be written as

\begin{equation}
\theta_i^{(n)}=G_i(\vec{\theta}^{(n-1)}),\quad i=1\dots N;
\end{equation}
for the sake of illustration, let us discuss it for the UBCM case. In this particular case, the set of equations above can be rewritten as

\begin{equation}
\theta_i^{(n)}=-\ln\left[\frac{k_i(\mathbf{A}^*)}{\sum_{\substack{j=1\\(j\neq i)}}^N\left(\frac{e^{-\theta_j^{(n-1)}}}{1+e^{-\theta_i^{(n-1)}-\theta_j^{(n-1)}}}\right)}\right],\quad i=1\dots N.
\end{equation}

Since all components of the map $\mathbf{G}$ are continuous on $\mathbb{R}^N$, the map itself is continuous on $\mathbb{R}^N$. Hence, a fixed point exists. Let us now consider its Jacobian matrix and check the magnitude of its elements. In the UBCM case, one finds that

\begin{eqnarray}
\frac{\partial G_i}{\partial\theta_i}&=&\frac{\sum_{\substack{j=1\\(j\neq i)}}^N\frac{e^{-\theta_i-2\theta_j}}{\left(1+e^{-\theta_i-\theta_j}\right)^2}}{\sum_{\substack{j=1\\(j\neq i)}}^N\left(\frac{e^{-\theta_j}}{1+e^{-\theta_i-\theta_j}}\right)}=\frac{\sum_{\substack{j=1\\(j\neq i)}}^N\left(\frac{e^{-\theta_i-\theta_j}}{1+e^{-\theta_i-\theta_j}}\right)^2}{\sum_{\substack{j=1\\(j\neq i)}}^N\left(\frac{e^{-\theta_i-\theta_j}}{1+e^{-\theta_i-\theta_j}}\right)}\nonumber\\
&=&1-\frac{\text{Var}[k_i]}{\langle k_i\rangle}=1-\frac{\sum_{j\neq i}\text{Cov}[k_i,k_j]}{\langle k_i\rangle},\quad\forall\:i\nonumber\\
\end{eqnarray}
and

\begin{eqnarray}
\frac{\partial G_i}{\partial\theta_j}&=&-\frac{\frac{e^{-\theta_j}}{\left(1+e^{-\theta_i-\theta_j}\right)^2}}{\sum_{\substack{j=1\\(j\neq i)}}^N\left(\frac{e^{-\theta_j}}{1+e^{-\theta_i-\theta_j}}\right)}=-\frac{\frac{e^{-\theta_i-\theta_j}}{\left(1+e^{-\theta_i-\theta_j}\right)^2}}{\sum_{\substack{j=1\\(j\neq i)}}^N\left(\frac{e^{-\theta_i-\theta_j}}{1+e^{-\theta_i-\theta_j}}\right)}\nonumber\\
&=&-\frac{\text{Cov}[k_i,k_j]}{\langle k_i\rangle},\quad\forall\:i,j.
\end{eqnarray}

Let us notice that 1) each element of the Jacobian matrix is a continuous function $\mathbb{R}^N\rightarrow\mathbb{R}$ and that 2) the following relationships hold

\begin{equation}
\left|\frac{\partial G_i}{\partial\theta_j}\right|\leq1,\quad\forall\:i,j;
\end{equation}
unfortunately, however, when multivariate functions are considered, the set of conditions above is not enough to ensure convergence to the fixed point for \textit{any} choice of the initial value of the parameters. What is needed to be checked is the condition $||J_\mathbf{G}(\vec{\theta})||<1$, with $J$ indicating the Jacobian of the map (i.e. the matrix of the first, partial derivatives above) and $||.||$ any natural matrix norm: the validity of such a condition has been numerically verified case by case.

\end{document}